\documentclass{elsart}
\usepackage{amsfonts}
\usepackage{amsmath}
\usepackage{graphicx}
\usepackage{amssymb}
\newtheorem{theorem}{Theorem}
\newtheorem{definition}[theorem]{Definition}
\newtheorem{remark}[theorem]{Remark}
\newenvironment{proof}[1][Proof]{\noindent\textbf{#1.} }{\ \rule{0.5em}{0.5em}}
\begin{document}
\begin{frontmatter}
\title{A Continuum Theory for Unstructured Mesh Generation in Two Dimensions}
\author{Guy Bunin}
\address{Department of Physics, Technion,\\
Haifa 32000, Israel\\
buning@tx.technion.ac.il}
\begin{abstract}
A continuum description of unstructured meshes in two dimensions, both for
planar and curved surface domains, is proposed. The meshes described are those
which, in the limit of an increasingly finer mesh (smaller cells), and away
from irregular vertices, have ideally-shaped cells (squares or equilateral
triangles), and can therefore be completely described by two local properties:
local cell size and local edge directions.  The connection between the two 
properties is derived by defining a Riemannian manifold whose geodesics trace 
the edges of the mesh.
A function $\phi$, proportional to the logarithm of the cell size, is shown to obey the
Poisson equation, with localized charges corresponding to irregular vertices.
The problem of finding a suitable manifold for a given domain is thus shown to
\emph{exactly} reduce to an Inverse Poisson problem on $\phi$, of finding a
distribution of localized charges adhering to the conditions derived for boundary
alignment. Possible applications to mesh generation are discussed.
\end{abstract}
\begin{keyword}
Unstructured mesh generation, differential geometry.
\PACS
\end{keyword}
\end{frontmatter}

\section{Overview\label{sec:overview}}

A mesh is a partition of a domain into smaller parts, typically with simpler
geometry, called cells. In two dimensions, both on the plane and on curved
surfaces, cells are usually triangles or quadrilaterals. The shapes of the
cells may be important; for many applications, cells with shapes similar to an
equilateral triangle or a square are preferred. The problem of mesh generation
can then be seen as an optimization problem:\ to find a partition of a domain
into well-shaped cells, possibly under additional demands, such as cell size requirements.

The mesh generation problem has been the subject of extensive research. Many
techniques for creating meshes exist, especially in two dimensions
\cite{owen},\cite{Thompson},\cite{alliez2005}. Nevertheless, some of the most
popular techniques, which create good meshes in many cases, are heuristic in
nature, and may create less than optimal meshes for some inputs. Two of the
inherent characteristics of the mesh generation problem seem to make it very
difficult to solve:

\begin{description}
\item[(Characteristic 1)] The constraints on cells' shapes are \emph{global}.
That is, the shape of one cell is constrained by the possible shapes of its
neighbors, which in turn are constrained by the shapes of their neighbors, and
so on. Thus, at least in principle, the constraints on the mesh layout extend
over the whole domain (or more precisely, over each connected component of the domain).

\item[(Characteristic 2)] The problem combines \emph{discrete and continuous
aspects}. The number of cells and the mesh connectivity (i.e.: which cells are
neighbors? which faces do they share?) are of discrete nature, whereas the
locations of the vertices can vary continuously. These aspects are closely
intertwined, preventing the sole use of purely discrete techniques (algebraic,
graph-theoretic, etc.), or techniques designed for use in problems of
continuous nature.
\end{description}

\begin{figure}
[ptb]
\begin{center}
\includegraphics[
trim=0.439322in 0.000000in 0.000000in 0.000000in,
height=2.0764in,
width=2.3506in
]
{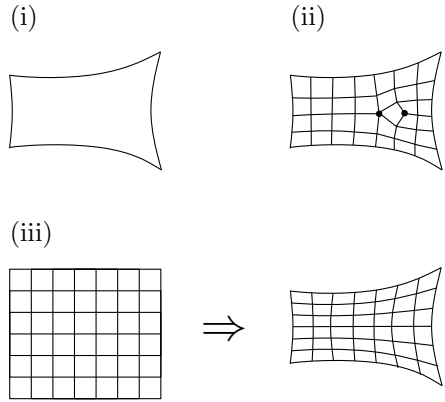}
\caption{Unstructured vs. structured meshes. (i) Input domain. (ii)
Unstructured mesh. Irregular vertices are marked. (iii) Structured mesh
created by mapping a regular grid.}
\label{mapping_vs_unstructured}
\end{center}
\end{figure}

The meshes created by mesh generation algorithms\ can be divided into
structured meshes, and unstructured meshes. A structured mesh is a mesh whose
connectivity is that of a regular grid, see Fig. \ref{mapping_vs_unstructured}
,(iii). By assuming the connectivity of the mesh beforehand, the problem of
creating such a mesh is considerably simplified (see Characteristic 2 above),
and reduces to the problem of assigning locations to the mesh vertices. One
way of doing that is by finding a \emph{mapping function} that maps the domain
of the regular grid to the domain to be meshed, and using it to map the
vertices. Such techniques are known as \emph{mapping techniques}. For small
enough cells, the differential properties of the mapping function at the
cell's location dictate its shape. If, for example, a mapping is angle
preserving, then the inner angle of a small enough cell will be approximately
preserved under the mapping. For a survey, see Ref. 2. Mapping techniques have
also been offered for unstructured meshes, (in the context of smoothing a
given mesh see \cite{hansen},\cite{Liseikin} and references therein), as long
as the connectivity of the mesh is given beforehand.

\begin{figure}
[tbh]
\begin{center}
\includegraphics[
trim=0.000000in 0.333892in 0.000000in 0.196209in,
height=1.9052in,
width=4.5558in
]
{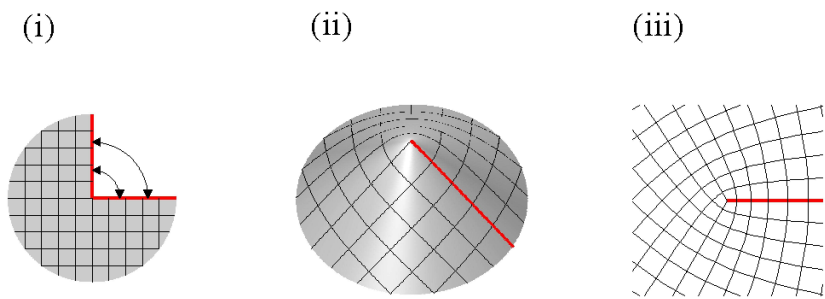}
\caption{Cone point. (i) The Volterra construction. (ii) A cone created by the
Volterra construction. (iii) The cone mapped onto the plane. Alternatively:
geodesics on a manifold containing a conical singularity.}
\label{3_cone_all}
\end{center}
\end{figure}

Just as a structured mesh can be imagined as created by mapping of a region of
the plane to the domain to be mapped, an unstructured mesh in two dimensions
can be imagined as a surface, that is mapped onto the domain to be meshed. The
simplest example is that of an unstructured mesh with just one irregular
vertex (a vertex that has more or less than four cells incident upon it). The
surface to be mapped in this case is a cone. This can be visualized using the
Volterra construction \cite{volterra}: Consider a piece of paper with an
angular section cut out, see Fig. \ref{3_cone_all},(i). If the two edges of
the section are identified, i.e. glued together, the paper will assume the
shape of a cone, see Fig. \ref{3_cone_all},(ii). If the cone is then mapped to
the plane, a regular grid drawn on the cone would be mapped to an unstructured
mesh such as the one shown in Fig. \ref{3_cone_all},(iii). The mapping shown
in Fig. \ref{3_cone_all},(iii) has the special property of being
\emph{conformal}: a small square on the cone is approximately mapped to a
square on plane. This creates well shaped cells in the resulting mesh: in the
limit of an increasingly smaller cell, its square shape is preserved under the
conformal mapping. A similar construction can be imagined for creating an
unstructured mesh with more than one irregular vertex; each irregular vertex
will then correspond to one \textquotedblleft cone tip\textquotedblright\ of
the surface.

The approach of the present work to the problem of creating unstructured
meshes can be expressed as follows: given a domain to be meshed,
\emph{what\ surface,\ with\ a\ square\ grid\ drawn\ upon\ it,
can\ be\ mapped\ conformally into this domain? }Thus, it is not just the
mapping function that is sought after, but rather the surface to be mapped
together with the mapping function. Unlike mapping techniques, however, the
mapping function is required to be conformal.

A mathematical framework highly suitable for dealing with such questions is
Riemannian geometry. Given the domain to be meshed, Riemannian geometry allows
one to define a \textquotedblleft new geometry\textquotedblright\ for that
domain. This includes a redefinition of the distances between points of the
domain, and it is used here to redefine distances such that \emph{a cell edge
have unit length}. Thus, instead of defining the surface that is mapped and
the mapping function separately, both are treated together, as the mapping
induces a new distance definition on the domain to be meshed. For example, the
cone in Fig. \ref{3_cone_all},(ii) and the surface in Fig. \ref{3_cone_all}
,(iii) have the \textquotedblleft same geometry\textquotedblright\ (i.e. are
isometric) if the distances in Fig. \ref{3_cone_all},(iii) are \emph{defined}
such that cell-edges have unit length. Using the terminology of Riemannian
geometry, the problem can be restated as assigning a new metric to the domain
being meshed, having the following properties:

\begin{description}
\item[(Property 1)] The metric is locally \emph{flat} everywhere, except at
some points, called \emph{cone points}. A \textquotedblleft
flat\textquotedblright\ region can be imagined as a bent, but not stretched,
piece of paper. The cone points are the \textquotedblleft tips of the
cones\textquotedblright\ as described above.

\item[(Property 2)] A single real function $\phi$ is defined in the domain
(except at cone points), such that at any given point $p$, the new metric
$\tilde{g}_{ij}\left(  p\right)  $ at $p$ is proportional to the original
metric $g_{ij}\left(  p\right)  $: $\tilde{g}_{ij}\left(  p\right)
=e^{2\phi\left(  p\right)  }g_{ij}\left(  p\right)  $, with $1\leq i,j\leq2$,.
(On the plane with Cartesian coordinates $g_{ij}=\delta_{ij}$.) We stress that
the proportionality $e^{2\phi}$ can vary throughout the domain. The quantity
$e^{-\phi}$ is the local change in the distance definition and will be
associated with the local cell size. (In manifold theory terminology, the
metric $\tilde{g}_{ij}$ is \emph{conformally}\textit{ }\emph{related} to
$g_{ij}$).
\end{description}

\begin{figure}
[tbh]
\begin{center}
\includegraphics[
trim=0.000000in 0.108093in 0.000000in 0.118187in,
height=1.1139in,
width=1.932in
]
{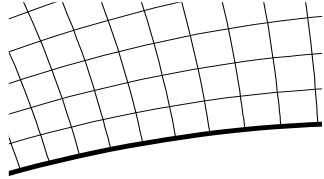}
\caption{Boundary alignment. Boundary marked with heavy line.}
\label{boundary_aligment}
\end{center}
\end{figure}

These two properties are not sufficient for our purposes. First of all, as the
Volterra construction implies, for a grid of squares to be drawn around a cone
tip, the total angle as measured around the tip must be a multiple of $\pi/2$
radians. Secondly, the mesh must be aligned along the boundary, see Fig.
\ref{boundary_aligment}. To formalize these demands, we define the direction
of mesh-edges at each point on the surface. Since the edges are assumed to
form right angles at incidence, this direction is defined up to an addition of
$\pi/2$ radians\footnote{In the case of a triangular mesh, the direction will
be defined up to an addition of $\pi/3$.}. The four directions at each point
will hence be called a \emph{cross}, and the field of directions on the entire
domain being meshed will be called a \emph{cross-field}. The cross-field is
related to the new metric by requiring that the curves, generated by following
the directions of the cross-field, along which the edges will be laid, will be
geodesics of the new metric $\tilde{g}_{ij}$. (Geodesics are the
generalizations of straight lines for surfaces and manifolds.) The boundary
alignment is formalized by requiring that the crosses be aligned with the boundary.

We therefore add the following property to the required properties of the new metric:

\begin{description}
\item[(Property 3)] A cross-field exists. (Exact definition is given in
section \ref{sec:cross_field}.)
\end{description}

Property 1 states that the Gaussian curvature $\tilde{K}$ of the manifold with
metric $\tilde{g}_{ij}$ be identically zero everywhere, except at cone points.
Combined with the equation $\tilde{g}_{ij}=e^{2\phi}g_{ij}$ of Property 2, the
two formulas give a remarkably simple result, viz. that $\phi$ must obey the
Poisson equation $\nabla^{2}\phi=K$ everywhere exept at cone points, where $K$
is the Gaussian curvature of the surface. This is a well known result in
conformal geometry, see section \ref{sec:cross_field}.

The main result of the present work is that for a cross-field to exist the
function $\phi$, which uniquely defines the manifold, has to obey the Poisson
equation $\nabla^{2}\phi=K+\rho$, with $K$ the Gaussian curvature of the
surface to be meshed ($K$ vanishes for the plane), and $\rho$ a sum of
\textquotedblleft localized charges\textquotedblright\ (Dirac delta
functions). The charges are placed at the cone points' locations, thus
corresponding to the irregular vertices of the mesh. The charge strength is
equal to the cone excess angle (the difference between the cone angle and
$2\pi$) which corresponds to the number of cells incident on the irregular
vertex. For example, in the manifold shown in Fig. \ref{3_cone_all},(iii), the
cone point has an excess angle of $-\pi/2$. The charge of the cone point,
located at a point $p$, corresponding to such a excess angle is $-\frac{\pi
}{2}\delta_{p}^{\left(  2\right)  }$, where $\delta_{p}^{\left(  2\right)  }$
is the Dirac delta function in two dimensions. This is the charge of any
singularity corresponding to an irregular vertex surrounded by three cells.
Along with the boundary alignment conditions on $\phi$, the problem of finding
an appropriate manifold is thus reduced to an \emph{Inverse Poisson} problem,
of finding a charge distribution adhering to these conditions. The reduction
gives \emph{exact, global} relations that the function $\phi$ must obey.

The inverse Poisson problem on planar domains is of interest in many fields of
science and engineering (see references in section \ref{sec:case_C}); the
relevance of existing techniques to the present application remains to be
examined. Another possible application is to the problem of creating a surface
mesh aligned with predefined directions, which has recently attracted much
attention \cite{alliez},\cite{shimada},\cite{ray},\cite{tong}. The algorithm
described in \cite{ray} creates conformal parametrizations and meshes
approximately aligned with predefined directions. In that work, the local size
and the local direction are treated as independent variables. In order to
reduce the number of singularities, a preprocessing step that modifies the
cell size demand (the \textquotedblleft curl correction\textquotedblright
\ process) can be applied. In contrast, the theory developed in the present
work uses conformality to a-priory link the cell-size and cell-direction
fields, leaving just one field to work with, either the cell-size or the
cell-direction. The applications of this theory to surface parameterization
and surface quadrangulation problems are an interesting subject for future work.

Conformal parametrization of manifolds with conical singularities have been
used in surface parametrization problems. For reviews and related work, see
\cite{floater}-\cite{ray}. Mesh generation with boundaries and surface
parametrization are different problems, because of the boundary alignment
requirement in mesh generation. For example, the parametrization problem for
planar domains is trivial: the coordinates of the plane form a good
parametrization, but do not solve the mesh generation problem.

The rest of the article is organized as follows: In section
\ref{sec:parallel_transport_and_geodesics} some elements of differential
geometry are shortly reviewed. In section \ref{sec:cross_field} the
cross-field and $\phi$-manifold are defined. Section
\ref{sec:relation_to_mesh_generation} discusses the relation of the
definitions in previous sections to mesh generation. Cone points are analyzed
in section \ref{sec:cone_points}. The necessary and sufficient conditions for
cross-field existence are developed in section
\ref{sec:boundary_alignment_conditions}. The set of conditions derived in
sections \ref{sec:cone_points},\ref{sec:boundary_alignment_conditions} forms
the core result of the work. In section \ref{sec:discussion} the theory is
discussed through four case studies. Case Studies A,B are used to discuss the
meaning of the conditions derived before. In Case Study C the possible
structure of a mesh generation algorithm is discussed. An example
quadrilateral mesh problem, though admittedly simple and artificial,
demonstrates\ how, given an input, a manifold with cone points is constructed,
adhering to the imposed conditions. This manifold allows the construction of
meshes with irregular vertices; at any given point in the domain that is not
an irregular vertex, at the limit of increasingly finer meshes, the cells'
shapes tend to a square. Case study D gives an example of a curved surface
meshing problem, and how it is solved.

\section{Parallel Transport and
Geodesics\label{sec:parallel_transport_and_geodesics}}

In this section basic facts from differential geometry, required in subsequent
sections, are shortly reviewed. More complete accounts can be found in any
textbook on differential geometry, such as \cite{millman},\cite{laugwitz}.

For a surface $D$ embedded in three-dimensional space, distances and angles on
the surface can be defined by the embedding. In some coordinate system, denote
by $g_{ij}$ the metric given by the embedding. For example, on the plane with
Cartesian coordinates, $g_{ij}=\delta_{ij}$, the Kronecker delta. Another
metric, $\tilde{g}_{ij}$, is said to be \emph{conformally related} to $g_{ij}
$, if there exists a real function $\phi$ on the surface such that
\begin{equation}
\tilde{g}_{ij}=e^{2\phi}g_{ij}. \label{eq:conformal_relation}
\end{equation}

Given two vectors $\mathbf{x}=(x^{1},x^{2}),\mathbf{y}=(y^{1},y^{2})$ defined
at some point on a surface or Riemannian manifold, the angle between them is
given by
\begin{equation}
\cos\theta=(x^{i}g_{ij}y^{j})/(\sqrt{x^{k}g_{kl}x^{l}}\sqrt{y^{m}g_{mn}y^{n}
}), \label{eq:cos_theta}
\end{equation}
where summation over repeated indices (the Einstein convension) is assumed.
The angle $\cos\tilde{\theta}$ as measured with the metric $\tilde{g}_{ij}$ is
\begin{equation}
\cos\tilde{\theta}=(x^{i}\tilde{g}_{ij}y^{j})/(\sqrt{x^{k}\tilde{g}_{kl}x^{l}
}\sqrt{y^{m}\tilde{g}_{mn}y^{n}}). \label{eq:cos_theta_tilde}
\end{equation}
Substituting $\tilde{g}_{ij}=e^{2\phi}g_{ij}$ in Eq. (\ref{eq:cos_theta_tilde}
), and comparing with Eq. (\ref{eq:cos_theta}) it is found that $\cos
\tilde{\theta}=\cos\theta$,\ so the measurement of angles using the two
conformally related metrics agrees.

\begin{figure}
[ptb]
\begin{center}
\includegraphics[
height=0.9556in,
width=2.3194in
]
{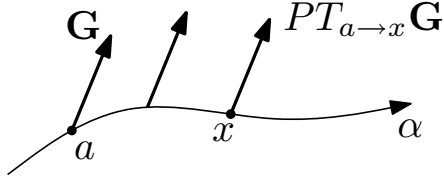}
\caption{Parallel transport of a vector $\mathbf{G}$ along a curve $\alpha$.}
\label{parallel_transport_basic}
\end{center}
\end{figure}

Given a vector field on the plane (with the Euclidean metric) the question of
whether two vectors at two different points are parallel has a definite
answer. This is not the case on curved surfaces, and more generally, for
Riemannian manifolds. There, the definition of parallel vectors at different
locations generally depends on the path chosen between the points. For a curve
$\alpha$ connecting points $a$ and $b$, the parallel transport of a vector
from $a$ to $b$ along $\alpha$ can be defined. If $\mathbf{G}$ is such a
vector at $a$ we denote its parallel translate to $b$ along $\alpha$ by
$PT_{a\overset{\alpha}{\longrightarrow}b}\mathbf{G}$, see Fig.
\ref{parallel_transport_basic}.

\begin{figure}
[ptb]
\begin{center}
\includegraphics[
height=1.1606in,
width=2.4742in
]
{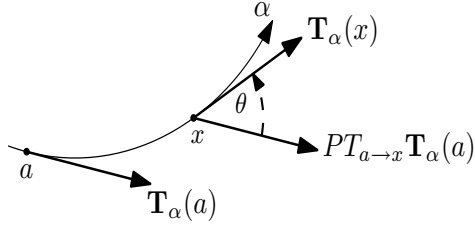}
\caption{Definition of $\kappa_{g}$.}
\label{parallel_transport}
\end{center}
\end{figure}

The geodesic curvature $\kappa_{g}$ of a curve is the amount by which a curve
\textquotedblleft turns\textquotedblright. On the plane (with the Euclidean
metric) turning is measured by the change of the angle of the tangent vector
to the curve. On a surface (and, more generally, on a Riemannian manifold) the
angle is defined relative to a vector that is parallel translated along the
very same curve. For a curve $\alpha$, denote the tangent vector at $x$ by
$\mathbf{T}_{\alpha}\left(  x\right)  $. Define $\theta\left(  x\right)
=\measuredangle\left(  PT_{a\overset{\alpha}{\longrightarrow}x}\mathbf{T}
_{\alpha}\left(  a\right)  ,\mathbf{T}_{\alpha}\left(  x\right)  \right)  $.
Then
\begin{equation}
\kappa_{g}=\frac{d\theta}{ds}, \label{kappa_is_dtheta_ds}
\end{equation}
where $s$ is the length parameterization of $\alpha$, see Fig.
\ref{parallel_transport}.

The parallel transport depends on the metric, so a curve can have different
geodesic curvatures under different metrics. Let $\kappa_{g},\tilde{\kappa
}_{g}$ be the geodesic curvatures of a curve $\alpha$ at some point $x$, with
the metrics $g_{ij},\tilde{g}_{ij}$ respectively, related as in Eq.
(\ref{eq:conformal_relation}). (Henceforth, all quantities relating to the
metric $\tilde{g}_{ij}$ will be marked with a tilde.) Then $\kappa_{g}
,\tilde{\kappa}_{g}$ are related by
\begin{equation}
\tilde{\kappa}_{g}=e^{-\phi}\left(  \kappa_{g}-\partial_{n}\phi\right)  ,
\label{kappa_g_relation}
\end{equation}
where $\partial_{n}\phi\equiv\partial\phi/\partial n$ is the derivative of
$\phi$ along the normal vector $\mathbf{N}_{\alpha}$, which is defined such
that $\left(  \mathbf{T}_{\alpha}\mathbf{,N}_{\alpha}\right)  $ form a right
hand system. Eq. (\ref{kappa_g_relation}) is derived in appendix
\ref{app:k_g_different_metrics}.

A geodesic is a generalization of a straight line. It is the curve whose
geodesic curvature vanishes. The equation for a geodesic of the metric
$\tilde{g}_{ij}$ is found by substituting $\tilde{\kappa}_{g}=0$ \ in Eq.
(\ref{kappa_g_relation}):
\begin{equation}
\kappa_{g}=\partial_{n}\phi\text{.} \label{phi_manifold_geodesic}
\end{equation}

Two integral theorems are used throughout the paper. The first is the
Gauss-Bonnet theorem, which relates the change in a vector undergoing parallel
transport along a closed loop, to the total Gaussian curvature inside the
loop. Let $\alpha\left(  s\right)  $ be a closed curve, $a\leq s\leq b,$
$\alpha\left(  a\right)  =\alpha\left(  b\right)  $, enclosing a region $R$.
The junction angles, $\theta_{1}..\theta_{N}$, measure the change in direction
of the tangent at junction points. As a convention we assume that $\alpha$ is
traversed in a counter-clockwise direction. let $\mathbf{V}$ be a vector at
$\alpha\left(  a\right)  $. Then the angle $\measuredangle\left(
\mathbf{V},PT_{a\overset{\alpha}{\longrightarrow}a}\mathbf{V}\right)  $ is
equal to
\begin{equation}
\measuredangle\left(  \mathbf{V},PT_{a\overset{\alpha}{\longrightarrow}
a}\mathbf{V}\right)  =2\pi-
{\displaystyle\oint\limits_{\mathbf{\alpha}}}
\kappa_{g}ds-\sum_{i=1}^{N}\theta_{i}=\int\int_{R}Kda \label{gauss_bonnet}
\end{equation}
where $K$ is the Gaussian curvature.

The second integral theorem is Green's theorem, also known as the divergence
theorem, or Gauss' theorem. Suppose $\alpha\left(  s\right)  $ is a curve
enclosing a region $R$, transversed in a counter-clockwise manner, and $\phi$
a function defined on the surface.\ Then
\begin{equation}
\int\int_{R}\nabla^{2}\phi da=-
{\displaystyle\oint\limits_{\mathbf{\alpha}}}
\partial_{n}\phi ds. \label{greens_theorem}
\end{equation}
The (unconventional) minus sign appears because the normal direction to
$\alpha$ was defined such that $\left(  \mathbf{T_{\alpha},N}_{\alpha}\right)
$ form a right hand system, so $\mathbf{N}_{\alpha}$ points \emph{inwards}.
For a surface the Laplacian operator $\nabla^{2}$ denotes the Laplace-Beltrami
operator \cite{laugwitz}.

If two metrics are conformally related as in Eq. (\ref{eq:conformal_relation}
), and the manifold with the metric $\tilde{g}_{ij}$ is flat, i.e.
\begin{equation}
\tilde{K}=0, \label{eq:flatness_assumption}
\end{equation}
a differential equation for $\phi$ can be derived. This can be done using the
expression for the curvature $\tilde{K}$ in terms of $\tilde{g}_{ij}$. It is
derived here in a different way, using the integral theorems quoted above, as
this technique is used again in subsequent sections.

Suppose that a region $R$ equipped with the metric $\tilde{g}_{ij}$ is flat,
i.e. Eq. (\ref{eq:flatness_assumption}) holds in $R$. Then according to the
Gauss-Bonnet theorem, Eq. (\ref{gauss_bonnet}),
\begin{equation}
{\displaystyle\oint\limits_{\mathbf{\alpha}}}
\tilde{\kappa}_{g}d\tilde{s}=2\pi-\sum_{i=1}^{N}\theta_{i}\text{.}
\label{derive1:lap_phi_eq_k}
\end{equation}
The length element for the two metrics discussed are related by
\begin{equation}
d\tilde{s}=\sqrt{dx^{i}\tilde{g}_{ij}dx^{j}}=e^{\phi}\sqrt{dx^{i}g_{ij}dx^{j}
}=e^{\phi}ds. \label{ds_vs_ds_tilde}
\end{equation}
Substitute Eq. (\ref{kappa_g_relation}),(\ref{ds_vs_ds_tilde}) in Eq.
(\ref{derive1:lap_phi_eq_k}) to get
\begin{equation}
{\displaystyle\oint\limits_{\mathbf{\alpha}}}
\tilde{\kappa}_{g}d\tilde{s}=
{\displaystyle\oint\limits_{\mathbf{\alpha}}}
e^{-\phi}\left(  \kappa_{g}-\partial_{n}\phi\right)  e^{\phi}ds=
{\displaystyle\oint\limits_{\mathbf{\alpha}}}
\kappa_{g}ds+\int\int_{R}\nabla^{2}\phi da\text{.}
\label{derive2:lap_phi_eq_k}
\end{equation}
Here Green's theorem, Eq. (\ref{greens_theorem}), was used. Subtracting Eq.
(\ref{derive1:lap_phi_eq_k}) from (\ref{derive2:lap_phi_eq_k}),
\begin{equation}
0=2\pi-\sum_{i=1}^{N}\theta_{i}-
{\displaystyle\oint\limits_{\mathbf{\alpha}}}
\kappa_{g}ds-\int\int_{R}\nabla^{2}\phi da=\int\int_{R}\left(  K-\nabla
^{2}\phi\right)  da\text{,}
\end{equation}
where Gauss-Bonnet was used. Since this result is true for an arbitrary flat
region $R$, the integrand $K-\nabla^{2}\phi$ must vanish, i.e., for any point
in $D$ where Eq. (\ref{eq:flatness_assumption}) holds:
\begin{equation}
\nabla^{2}\phi=K\text{.} \label{lap_phi_eq_K}
\end{equation}
Eq. (\ref{lap_phi_eq_K}) is a differential equation for $\phi$. It is a
well-known result in conformal geometry, see e.g. \cite{Chang},\cite{Aubin}.

\section{Cross-field and $\phi$-manifold definitions\label{sec:cross_field}}

The input to the mesh generation problem is assumed to be a surface $D$
embedded in three dimensional Euclidean space, such that:

\begin{description}
\item[(i)] Its boundary $\partial D$ is a union of a finite number of
connected components $\partial D=\cup\Gamma_{j}$.

\item[(ii)] Every component $\Gamma_{j}$ is a piecewise $C^{2}$ closed curve.
\end{description}

The points where a boundary component $\Gamma_{j}$ is not differentiable will
be called \emph{junction points}. The set of all junction points will be
denoted by $J$.

The edges of the final mesh are to be laid along geodesic curves of a manifold
that will be defined below, the $\phi$\emph{-manifold}. To create a high
quality mesh, these curves should cross each other, and reach the boundary, at
certain angles. In the case of a quadrilateral mesh, a cell with right inner
angles is preferred for many applications. In the case of a triangular mesh,
the preferred inner angle is $\pi/3$ radians. It is therefore natural to
define the direction of edges at a point to within an addition of $\pi/2$
radians in the quadrilateral case, and $\pi/3$ radians in the triangular case.
This difference between the two cases leads to slightly different results; for
clarity of presentation, the quadrilateral case is presented first, and the
results for the triangular case are defered to Appendix \ref{Triangular_mesh}.

The directions of edges at a point will be called a \emph{cross}. The field of
edge directions will be called a \emph{cross-field}.

\begin{definition}
[Cross]First, define an equivalence $\sim$ of vectors in $\mathbb{R}^{2}$: for
2 vectors $\mathbf{v}_{1},\mathbf{v}_{2}\in\mathbb{R}^{2}$, $\mathbf{v}
_{1}\sim\mathbf{v}_{2}$ if and only if $\mathbf{v}_{1}$,$\mathbf{v}_{2}$ are
parallel or perpendicular. A \emph{cross} is an element of $\mathbb{R}
^{2}/\sim$.
\end{definition}

Thus, a given cross is a set of vectors, every pair of which are either
perpendicular or parallel.

If two vectors, defined at some point, undergo parallel transport along the
same curve, the angle between two vectors is preserved: the angle before the
parallel transport is equal to the angle after the parallel transport
\cite{millman}. Therefore, the cross equivalence class structure is preserved,
and the parallel transport of crosses is well-defined.

Let $P$ be a finite set of points in $\bar{D}\equiv D\cup\partial D$. Denote
the metric on $D$ by $g_{ij}$. Supoose a metric $\tilde{g}_{ij}$ is defined on
$D\backslash P$. The following definition of a cross-field assures that the
flow lines of the cross-field are geodesics, and that the crosses on the
boundary are aligned with boundary.

\begin{definition}
[Cross-field]\label{cross_field_definition} A \emph{cross-field} on a given
manifold with metric $\tilde{g}_{ij}$ is a mapping $V:\bar{D}\backslash\left(
P\cup J\right)  \rightarrow\mathbb{R}^{2}/\sim$ such that:

\begin{description}
\item[(i)] For points $a,b\in D\backslash\left(  P\cup J\right)  $, the
parallel transport under the metric $\tilde{g}_{ij}$ of $V\left(  a\right)  $
to $b$ along a curve $\alpha$ is independent of $\alpha$, and is equal to
$V\left(  b\right)  $: $\widetilde{PT}_{a\rightarrow b}V\left(  a\right)
=V\left(  b\right)  $.

\item[(ii)] For $a\in$ $\Gamma_{j}\backslash\left(  J\cup P\right)  $, the
tangent belongs to the cross there: $\mathbf{T}_{\Gamma_{j}}\left(  a\right)
$ $\in V(a)$.
\end{description}
\end{definition}

\begin{figure}
[ptb]
\begin{center}
\includegraphics[
height=1.8421in,
width=2.2329in
]
{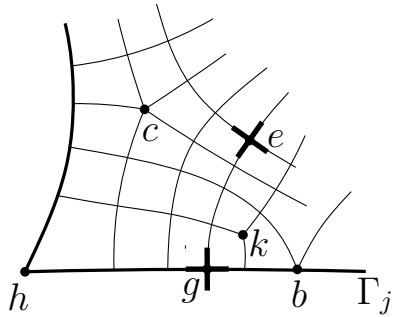}
\caption{Illustration of a cross-field, and its cross-field geodesics. The
cross-field is represented by the crosses. Thin lines represent the
cross-field geodesics. Heavy lines represent the boundary. Crosses at points
$g,e$ are shown. Points $c,k,b\ $are in $P$, hence no crosses are defined at
these points. $h$ is a junction point: $h\in J$.}
\label{c-frame_illustration}
\end{center}
\end{figure}

The flow-lines curves of a cross-field are geodesics of the metric $\tilde
{g}_{ij}$. That is, a geodesic aligned with the cross-field at one point
(i.e., whose tangent belongs to the cross at that point), is aligned with the
cross-field everywhere else on the curve. This is because both the cross-field
and the tangent to a geodesic are parallel-translated along the geodesic, see
definition \ref{cross_field_definition},(i), and the discussion preceding Eq.
(\ref{phi_manifold_geodesic}). Geodesic curves aligned with the cross-field
will be called \emph{cross-field geodesics}. Cross-field geodesics and their
relation to the cross-field are illustrated in Fig. \ref{c-frame_illustration}.

As before, let $P$ be a finite set of points in $\bar{D}$. We now define the
$\phi$\emph{-manifold}.

\begin{definition}
[$\phi$-manifold]\label{def:phi_manifold}A $\phi$\emph{-manifold} is a
Riemannian manifold defined on $D\backslash P$, with metric $\tilde{g}_{ij}$
such that:

\begin{description}
\item[(i)] $\tilde{g}_{ij}$ is conformally related to $g_{ij}$, i.e.
$\tilde{g}_{ij}=e^{2\phi}g_{ij}$ where $\phi$ is a real function on
$D\backslash P$.

\item[(ii)] The $\phi$-manifold is locally flat, i.e. its Gaussian curvature
tensor $\tilde{K}=0$ for all points in $D\backslash P$.

\item[(iii)] For every boundary point $a\in\Gamma_{j}$, $a\notin J\cup P$, the
limits $\lim_{r\rightarrow a}\phi\left(  r\right)  $ and $\lim_{r\rightarrow
a}\mathbf{\nabla}\phi\left(  r\right)  $ exist and are continuous along
$\Gamma_{j}$ at $a$.

\item[(iv)] A cross-field with the metric $\tilde{g}_{ij}$ exists.
\end{description}
\end{definition}

The points in $P$ will be called \emph{cone points}. This name is justified in
section \ref{sec:cone_points}, see also section \ref{sec:overview}. For now,
the points in $P$ are just points where $\phi$ is undefined.

In the following sections, the requirement that a $\phi$-manifold exists is
translated into conditions on the function $\phi$.

\section{Relation to mesh generation \label{sec:relation_to_mesh_generation}}

The cross-field formulates the demand that mesh cells have certain inner
angles. In order to have square-like shapes, the cells should further have
edges of similar lengths. This is where the conformal metric property of the
$\phi$-manifold comes into play.

Integrating over Eq. (\ref{ds_vs_ds_tilde}), the length of a curve $\alpha$ on
a $\phi$-manifold is given by
\begin{equation}
\tilde{s}\left(  \alpha\right)  =\int_{\alpha}d\tilde{s}=\int_{\alpha}e^{\phi
}ds\text{.} \label{eq:curve_manifold length}
\end{equation}

The area of a region of a manifold is given by \cite{millman},\cite{laugwitz}
\begin{equation}
\tilde{A}\left(  R\right)  \equiv\int\int_{R}\sqrt{\tilde{g}}dx^{1}dx^{2}
=\int\int_{R}e^{2\phi}\sqrt{g}dx^{1}dx^{2}\text{,} \label{manifold_area}
\end{equation}
where $g\equiv\det\left(  g_{ij}\right)  ,$ $\tilde{g}\equiv\det\left(
\tilde{g}_{ij}\right)  $. The tilde denotes, as before, a quantity with
respect to $\phi$-manifold metric $\tilde{g}_{ij}$.

The following claim is a consequence of the isometry of a flat manifold to the
Euclidean plane \cite{laugwitz}.\ It is the \textquotedblleft manifold
version\textquotedblright\ of the properties of a rectangle.

\begin{figure}
[ptb]
\begin{center}
\includegraphics[
height=1.6527in,
width=2.6247in
]
{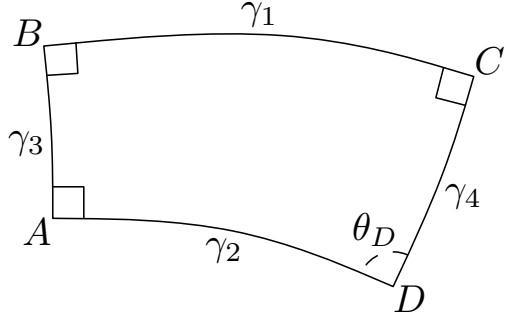}
\caption{A geodesic \textquotedblleft rectangle\textquotedblright.}
\label{rectangle}
\end{center}
\end{figure}

\begin{claim}
\label{claim:rectangle}Let $\gamma_{1},\gamma_{2},\gamma_{3},\gamma_{4}$ be 4
geodesic segments in a flat region of a manifold, organized as in Fig.
\ref{rectangle}. Suppose that the inner angles at the vertices $A,B,C$ are
right angles, and $\tilde{s}\left(  \gamma_{i}\right)  $ is the
manifold-length of the i-th side of the \textquotedblleft
rectangle\textquotedblright. Then the inner angle at $D$, $\theta_{D}$, is a
right angle, and $\tilde{s}\left(  \gamma_{1}\right)  =\tilde{s}\left(
\gamma_{2}\right)  ,\tilde{s}\left(  \gamma_{3}\right)  =\tilde{s}\left(
\gamma_{4}\right)  $. The area of the \textquotedblleft
rectangle\textquotedblright\ is\textbf{\ }$\tilde{s}\left(  \gamma_{1}\right)
\cdot\tilde{s}\left(  \gamma_{3}\right)  $.
\end{claim}

\bigskip

\begin{figure}
[ptb]
\begin{center}
\includegraphics[
height=1.4996in,
width=3.1306in
]
{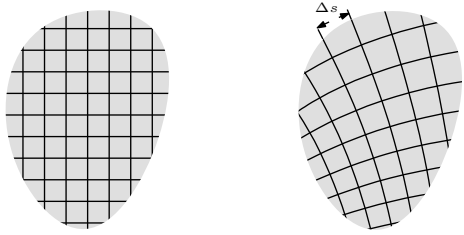}
\caption{A grid on an Euclidean plane (left), and on a flat manifold (right).}
\label{lay_grid}
\end{center}
\end{figure}

On an Euclidean plane, the properties of a rectangle allow one to lay a grid
on a region of the plane. A grid can be regarded as two families of mutually
perpendicular straight lines, with equal spacing between lines of each family.
The same can be done for a $\phi$-manifold using the \textquotedblleft
rectangle\textquotedblright\ properties stated in claim \ref{claim:rectangle},
see Fig. \ref{lay_grid}. The grid divides the space into square-like regions,
each bounded by four geodesic segments of manifold length $\Delta\tilde{s}$,
that will be called \emph{manifold-edges}. The regions enclosed by the
manifold-edges will be called \emph{manifold-cells}.

$\Delta\tilde{s}$ is a single, fixed number for the mesh. The overall size on
the surface or plane of the manifold-cells can be controlled by adding a
constant to $\phi$, so the freedom in choosing $\Delta\tilde{s}$ is redundant,
and we set $\Delta\tilde{s}=1$. According to Claim \ref{claim:rectangle} such
a cell has unit $\phi\,$-manifold area. According to Eq.
(\ref{eq:curve_manifold length}), for small enough manifold cells (large
enough $\phi$), a manifold edge of length $\Delta\tilde{s}=1$ has length
$\Delta s\left(  \mathbf{\gamma}_{e}\right)  \simeq e^{-\phi}$, and $e^{-\phi
}$ is interpreted as the \emph{local edge length}, or \emph{local cell size}.

\begin{figure}
[ptb]
\begin{center}
\includegraphics[
height=2.1811in,
width=2.8288in
]
{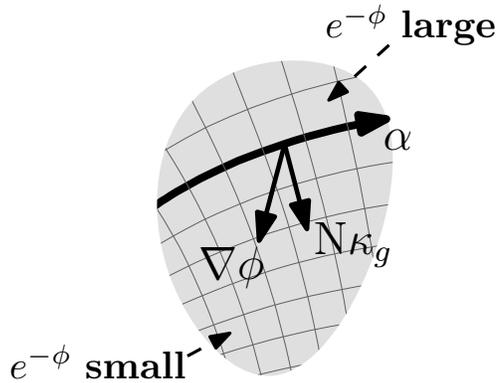}
\caption{Geometric interpretation of the formula $\kappa_{g}=\partial_{n}\phi
$: manifold-edges curve towards smaller cells. Cell size is proportional to
$e^{-\phi}$.}
\label{fig:lay_grid_interpretation}
\end{center}
\end{figure}

A geometrical interpretation of the relation$\ \kappa_{g}=\partial_{n}\phi$,
Eq. (\ref{phi_manifold_geodesic}), can now be given. For a mesh with
approximately square cells, the curving\ of manifold-edges is related to the
changes in cell size \emph{in the perpendicular direction}, see Fig.
\ref{fig:lay_grid_interpretation}. This is quantified in Eq.
(\ref{phi_manifold_geodesic}): $\kappa_{g}$ is the curvature of the lines
defining the edges, and $\partial_{n}\phi$ is the change in $\phi$, which is
related to local cell size by $e^{-\phi}$.

\section{Cone-points\label{sec:cone_points}}

\begin{figure}
[ptb]
\begin{center}
\includegraphics[
height=1.772in,
width=2.1499in
]
{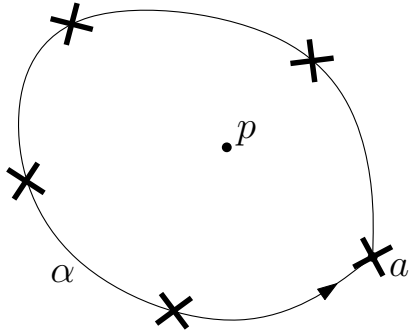}
\caption{Parallel transport around a cone point.}
\label{loop_around_charge}
\end{center}
\end{figure}

The existence of a cross-field restricts the possible behavior of $\phi$ in
the vicinity of cone points. Let $p\in P$ be a cone point, and $a\in
D\backslash P$ a point that is not a cone point. Let $\alpha$ be a simple
closed curve starting from $a$ that encloses $p$, and only $p$ of $P$, see
Fig. \ref{loop_around_charge}. The junction angles of $\alpha$ are denoted by
$\alpha_{1}..\alpha_{N}$. According to cross-field definition, the cross $V\left(
a\right)  $ is parallel translated to $V\left(  a\right)  $ along $\alpha$,
that is, a vector $\mathbf{y}\in V\left(  a\right)  $ is parallel translated
along $\alpha$, to $\mathbf{y}^{\prime}\in V\left(  a\right)  $. By the
defintion of a cross, $\mathbf{y,y}^{\prime}$ are either parallel or
perpendicular, and the angle $\measuredangle\left(  \mathbf{y},\mathbf{y}
^{\prime}\right)  $ is $k\pi/2$, $k\in\mathbb{Z}$. Using Eq.
(\ref{kappa_g_relation}),(\ref{gauss_bonnet}):
\begin{align}
k\pi/2  &  =\measuredangle\left(  \mathbf{y},\mathbf{y}^{\prime}\right)
=2\pi-\newline
{\displaystyle\oint\limits_{\mathbf{\alpha}}}
\tilde{\kappa}_{g}d\tilde{s}-\sum_{i=1}^{N}\alpha_{i}=\nonumber\\
&  =2\pi-
{\displaystyle\oint\limits_{\mathbf{\alpha}}}
\left(  \kappa_{g}-\partial_{n}\phi\right)  ds-\sum_{i=1}^{N}\alpha
_{i}\newline=
{\displaystyle\oint\limits_{\mathbf{\alpha}}}
\partial_{n}\phi ds+\int\int_{S}Kda\text{,} \label{charge_possible_values}
\end{align}
where $S$ is the region enclosed by $\alpha$.

On a $\phi$-manifold a manifold-cell has unit area, so in order to have a
finite number of cells, the the area of the $\phi$-manifold must be finite.
This will further restrict the type of singularity allowed at a point $p$, as
is now shown.

For a cone point $p$ on the surface let $U$ be a neighborhood of $p$ in which
there exist isothermal coordinates $\left(  x^{1},x^{2}\right)  $. In such
coordinates, which can always be found locally\footnote{The original proof,
due to Gauss, requires that the surface is analytic \cite{laugwitz}. There are
proofs with weaker assumptions, but these distinctions are immaterial for the
present purposes.}, the metric takes the form: $g_{ij}=F^{-1}\delta_{ij}$,
where $F\left(  x^{1},x^{2}\right)  $ is a real function of $\left(
x^{1},x^{2}\right)  $. Note that on the plane, the standard Cartesian
coordinates satisfy $g_{ij}=\delta_{ij}$, and hence are isothermal. In
isothermal coordinates the Laplace-Beltrami operator can be written
as\footnote{This can be seen by substituting the form of the metric tensor in
isothermal coordinates $g_{ik}=F^{-1}\delta_{ik}$ into the definition of the
Laplace-Beltrami operator: $\nabla^{2}\phi\equiv g^{ik}\phi,_{i;k}$, see e.g.
\cite{laugwitz}.}
\begin{equation}
{\nabla}^{2}\phi=F\left(  x^{1},x^{2}\right)  \left(  \frac{\partial^{2}\phi
}{\partial\left(  x^{1}\right)  ^{2}}+\frac{\partial^{2}\phi}{\partial\left(
x^{2}\right)  ^{2}}\right)  . \label{lap_in_isothermal}
\end{equation}
Eq. \ref{lap_in_isothermal} can be seen as a generalization of the usual
planar Laplacian, given by $F\left(  x^{1},x^{2}\right)  =1$. Let $B_{R}$ be a
disk of radius $R$\ in $\left(  x^{1},x^{2}\right)  $, i.e. the region for
which $\left(  x^{1}\right)  ^{2}+\left(  x^{2}\right)  ^{2}<R^{2}$. In
isothermal coordinates Eq. (\ref{lap_phi_eq_K}) reads
\[
\left(  \frac{\partial^{2}\phi}{\partial\left(  x^{1}\right)  ^{2}}
+\frac{\partial^{2}\phi}{\partial\left(  x^{2}\right)  ^{2}}\right)  =K\left(
x^{1},x^{2}\right)  /F\left(  x^{1},x^{2}\right)
\]
in the \emph{punctured} disk $B_{R}\backslash\left\{  p\right\}  $. In polar
isothermal coordinates $\left(  r,\psi\right)  $, (that is, the polar
coordinates corresponding to $\left(  x^{1},x^{2}\right)  $), the general form
of this solution can be written as (see e.g. \cite{tyn}):
\begin{equation}
\phi\left(  r,\psi\right)  =\phi_{P}+\frac{Q}{2\pi}\ln r+\sum_{n=1}^{\infty
}b_{n}r^{-n}\sin\left(  n\psi+c_{n}\right)  , \label{jackson_solution}
\end{equation}
where $\phi_{P}$ is a solution to the Poisson equation ${\nabla}^{2}\phi
_{P}=K/F$ in $B_{R}$ (\emph{including} $p$), and $Q,\left\{  b_{n}\right\}
_{n=1}^{\infty},\left\{  c_{n}\right\}  _{n=1}^{\infty}$ are all real numbers.

Let $\alpha$ be a curve tracing the circle $\left(  x^{1}\right)  ^{2}+\left(
x^{2}\right)  ^{2}=R^{2}$, in the counter-clockwise direction. For $\phi$
given by Eq. (\ref{jackson_solution}), the flux of $\mathbf{\nabla}\phi$
through $\alpha$ is
\begin{equation}
-\oint_{\alpha}\frac{\partial\phi}{\partial n}ds=\int\int_{B_{R}}
K/F\sqrt{\tilde{g}}dx^{1}dx^{2}+Q, \label{eq:flux_from_cone_poine}
\end{equation}
where Green's theorem, Eq. (\ref{greens_theorem}) was used to convert the
first term to a surface integral. In the limit $R\rightarrow0$ Eq.
(\ref{eq:flux_from_cone_poine}) becomes: $\oint_{\alpha}\frac{\partial\phi
}{\partial n}ds\rightarrow-Q$, and Eq. (\ref{charge_possible_values}) with
$S\rightarrow0$ becomes: $\oint_{\alpha}\frac{\partial\phi}{\partial
n}ds\rightarrow k^{\prime}\pi/2$ for some $k^{\prime}\in
\mathbb{Z}
$. Comparing the two limiting values for $\oint_{\alpha}\frac{\partial\phi
}{\partial n}ds$ we find
\begin{equation}
Q=k\frac{\pi}{2} \label{eq:charge_is_quantized}
\end{equation}
for some $k\in
\mathbb{Z}
$.

The quantity $Q$ will be called the \emph{charge} of the cone-point. Eq.
(\ref{eq:charge_is_quantized}) states that the charges must be multiples of
$\pi/2$.

In order to have a finite number of unit-area manifold cells, the
manifold-area of a neighbourhood of $p$ must be finite. The manifold area of
the disc $B_{R}$ of radius $R$ around $p$ is obtained by substituting Eq.
(\ref{jackson_solution}) into Eq. (\ref{manifold_area}):
\begin{align}
\tilde{A}\left(  B_{R}\right)   &  =\int\int_{B_{R}}e^{2\phi}\sqrt{g}
dx^{1}dx^{2}\nonumber\\
&  =\int_{0}^{R}\exp\left(  2\phi_{P}+\frac{k}{2}\ln r+2\sum_{n=1}^{\infty
}b_{n}r^{-n}\sin\left(  n\psi+c_{n}\right)  \right)  \sqrt{g}2\pi
rdr\nonumber\\
&  =2\pi\int_{0}^{R}r^{k/2+1}\exp\left(  2\sum_{n=1}^{\infty}b_{n}r^{-n}
\sin\left(  n\psi+c_{n}\right)  \right)  e^{2\phi_{P}}\sqrt{g}dr.
\label{eq: singularity_area1}
\end{align}
Since a solution of ${\nabla}^{2}\phi_{P}=K$ is bounded in a compact
region,$\ e^{2\phi_{P}}$ does not effect the convergence of the integral, nor
does $\sqrt{g}$, which is bounded away from zero. To converge, it is required
that $b_{n}=0$ for every $n$, and that $k>-4$. This restricts $\phi$ to the
form
\begin{equation}
\phi\left(  r\right)  =\phi_{P}+\frac{k}{4}\ln r.
\label{eq: point source one source}
\end{equation}
with $k>-4$. This is a solution of the Poisson equation for one point-source
\begin{equation}
\nabla^{2}\phi=K+\frac{k\pi}{2}\delta_{p}^{\left(  2\right)  },
\label{one_defect_solution}
\end{equation}
in a neighborhood of $r=0$. For many cone points Eq.
(\ref{one_defect_solution}) becomes:\newline

\textbf{Condition 1:}

$\phi$\textit{\ obeys the equation}
\begin{equation}
\nabla^{2}\phi=K+\frac{\pi}{2}\sum_{i=1..N}k_{i}\delta_{p_{i}}^{\left(
2\right)  }, \label{eq:condition1}
\end{equation}
\textit{the Poisson equation with point sources (delta functions) }
$\delta_{p_{i}}^{\left(  2\right)  }$\textit{, with }$k_{i}\in
\mathbb{Z}
$\textit{, }$k_{i}>-4$\textit{.}

\begin{figure}
[ptb]
\begin{center}
\includegraphics[
height=1.881in,
width=4.7556in
]
{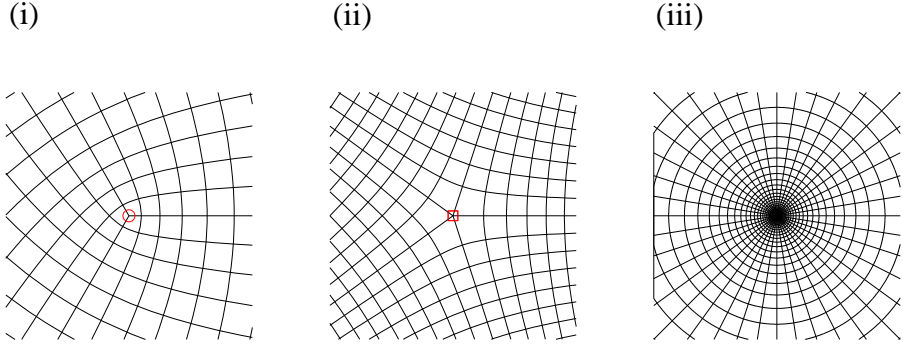}
\caption{Cross-field geodesics around a singularity. The geodesics are placed
at unit manifold-distances apart. (i) A $k=-1$ singularity. (ii) A $k=1$
singularity. (iii) A $k=-4$ singularity, with an infinite number of cells.}
\label{defects_3_5_inf}
\end{center}
\end{figure}

For a planar domain, Condition 1 can be rewritten as

\textbf{Condition 1 (planar domain):}

\textit{For }$r\in D\backslash P$\textit{, }$\phi\left(  r\right)
$\textit{\ can be written as}
\[
\phi\left(  r\right)  =\phi_{L}+\frac{1}{4}\sum_{i=1..N}k_{i}\ln\left(
\left\vert r-p_{i}\right\vert \right)  ,
\]
\textit{with }$k_{i}>-4$\textit{, }$P=\left\{  p_{i}\right\}  _{i=1..N}
$\textit{, and }$\nabla^{2}\phi_{L}=0$\textit{\ on }$D$\textit{.}

Fig. \ref{defects_3_5_inf} shows selected cross-field geodesics for a planar
domain around a singularity (with $\phi_{P}=0$), spaced at manifold-distances
of $\Delta\tilde{s}=1$\ from each other, for different singularity strengths.
In practice, good candidates for meshes will only have singularities of charge
$k\geq-1$, due to the inner-angles of cells incident on the singularity, see
also Remark \ref{star_geodesics} in section \ref{sec:case_C} below.

To summarize this section, it has been shown that the existence of a grid of
geodesics that follow a cross-field and create a finite number of cells,
restrict the form that the function $\phi$ can take. This is formulated in
Condition 1, stating that $\phi$ must obey the Poisson equation, with delta
function charges corresponding to cone points.

\section{Boundary Alignment
Conditions\label{sec:boundary_alignment_conditions}}

This section examines the conditions for boundary alignment of a cross-field,
see Definition \ref{cross_field_definition},(ii). Three conditions will shown
to be necessary. Sufficiency of the conditions is discussed in section
\ref{sec:sufficient_conditions}.

\begin{figure}
[ptb]
\begin{center}
\includegraphics[
height=1.8421in,
width=1.6786in
]
{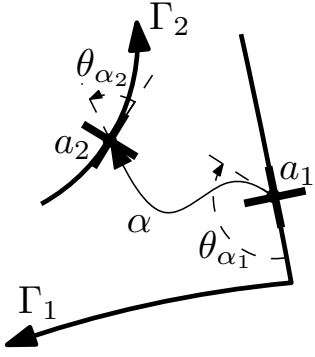}
\caption{Condition 4.}
\label{condition4_2_boundaries_surface}
\end{center}
\end{figure}

We start by developing an equation that will be used in the derivation of the
conditions below. Let $a_{1},a_{2}$ be two points on two boundary curves
$\Gamma_{1},\Gamma_{2}$, respectively. Note that $\Gamma_{1},\Gamma_{2}$ may
be the same boundary curve: $\Gamma_{1}=\Gamma_{2}$. Let $\alpha$ be some
curve from $a_{1}$ to $a_{2}$, see Fig. \ref{condition4_2_boundaries_surface}.
For $i=1,2$ denote by $\theta_{\alpha_{i}}=\measuredangle\left(
\mathbf{T}_{\Gamma_{i}}\left(  a_{i}\right)  ,\mathbf{T}_{\alpha}\left(
a_{i}\right)  \right)  $. By definition of a cross-field, the cross in $a_{1}$
must be parallel-translated along $\alpha$ to the cross in $a_{2}$. The cross
in $a_{1}$ contains $\mathbf{T}_{\Gamma_{1}}\left(  a_{1}\right)  $, and the
cross in $a_{2}$ contains $\mathbf{T}_{\Gamma_{2}}\left(  a_{2}\right)  $ so
\begin{equation}
n\frac{\pi}{2}=\measuredangle\left(  \widetilde{PT}_{a\overset{a}
{_{1}\longrightarrow}a_{2}}\mathbf{T}_{\Gamma_{1}}\left(  a_{1}\right)
,\mathbf{T}_{\Gamma_{2}}\left(  a_{2}\right)  \right)  =\eta+\theta
_{\alpha_{2}}\text{.} \label{eq:cond_4_proof_0}
\end{equation}
where $\eta$ is defined as
\begin{align}
\eta &  \equiv\measuredangle\left(  \widetilde{PT}_{a\overset{a}
{_{1}\longrightarrow}a_{2}}\mathbf{T}_{\Gamma_{1}}\left(  a_{1}\right)
,\mathbf{T}_{\alpha}\left(  a_{2}\right)  \right) \nonumber\\
&  =\measuredangle\left(  \widetilde{PT}_{a\overset{a}{_{1}\longrightarrow
}a_{2}}\mathbf{T}_{\Gamma_{1}}\left(  a_{1}\right)  ,\widetilde{PT}
_{a\overset{a}{_{1}\longrightarrow}a_{2}}\mathbf{T}_{\alpha}\left(
a_{1}\right)  \right)  +\measuredangle\left(  \widetilde{PT}_{a\overset
{a}{_{1}\longrightarrow}a_{2}}\mathbf{T}_{\alpha}\left(  a_{1}\right)
,\mathbf{T}_{\alpha}\left(  a_{2}\right)  \right) \nonumber\\
&  =\theta_{a_{1}}+\int_{a_{1}}^{a_{2}}\tilde{\kappa}_{g}d\tilde{s}\text{.}
\label{eq:cond_4_proof_1}
\end{align}
The last equality is due to the preservation of angle in parallel transport,
together with Eq. (\ref{kappa_is_dtheta_ds}). Eq. (\ref{kappa_g_relation}) and
(\ref{ds_vs_ds_tilde}) can now be used with Eq. (\ref{eq:cond_4_proof_1})
\[
\eta=\theta_{a_{1}}+\int_{a_{1}}^{a_{2}}\left(  \kappa_{g}-\partial_{n}
\phi\right)  ds\text{.}
\]
Substituting this into Eq. (\ref{eq:cond_4_proof_0}) we find the relation
\begin{equation}
\theta_{a_{2}}-\theta_{a_{1}}=\int_{a_{1}}^{a_{2}}\left(  \kappa_{g}
-\partial_{n}\phi\right)  ds+n\frac{\pi}{2}\text{.}
\label{eq:general_boundary_cond_equation}
\end{equation}
Eq. (\ref{eq:general_boundary_cond_equation}) is used below to develop the
next three conditions.

\begin{figure}
[ptb]
\begin{center}
\includegraphics[
height=1.0845in,
width=1.5454in
]
{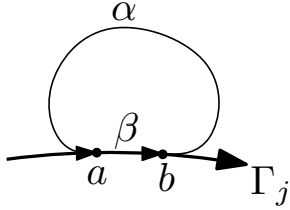}
\caption{Condition 2.}
\label{kappa_eq_d_phi_ds}
\end{center}
\end{figure}

Let $\beta$ be a segment of a boundary curve $\Gamma_{j}$, between points
$a,b\in\Gamma_{j}$ that are not junction points: $a,b\notin\left(  J\cup
P\right)  $, and suppose $\beta$ does not contain junction points. Choose some
smooth curve $\alpha$ from $b$ to $a$, whose tangents at $a,b$ coincide with
the tangents to $\beta$, see Fig. \ref{kappa_eq_d_phi_ds}, i.e. $\theta
_{a_{2}}=\theta_{a_{1}}=0$, and such that the region enclosed by $\alpha
,\beta$ does not include any cone points. According to Eq.
(\ref{eq:general_boundary_cond_equation}),
\begin{equation}
n\frac{\pi}{2}=\int_{\alpha}\partial_{n}\phi ds-\int_{\alpha}\kappa_{g}ds.
\label{cond_2_proof1_1}
\end{equation}
According to Gauss-Bonnet, Eq. (\ref{gauss_bonnet}),
\begin{equation}
\int_{\alpha}\kappa_{g}ds=2\pi-\int_{\beta}\kappa_{g}ds-\int\int_{S}Kda,
\label{cond_2_proof2}
\end{equation}
where $S$ is the area enclosed by $\alpha,\beta$. Green's theorem gives
\begin{equation}
\int_{\alpha}\partial_{n}\phi ds=-\int\int_{S}\nabla^{2}\phi da-\int_{\beta
}\partial_{n}\phi ds. \label{cond2_proof3}
\end{equation}
Substituting Eq. (\ref{cond_2_proof2}),(\ref{cond2_proof3}) and recalling that
$\nabla^{2}\phi=K$\ in $D\backslash P$ (see Eq. (\ref{lap_phi_eq_K})), Eq.
(\ref{cond_2_proof1_1}) now reads
\begin{equation}
n\pi/2=-2\pi+\int_{\beta}\kappa_{g}ds-\int_{\beta}\partial_{n}\phi ds.
\end{equation}

When $a\rightarrow b$, the integrals both tend to zero, and the equality is
possible only if $n\pi/2=-2\pi$. Then
\begin{equation}
\int_{\beta}\left(  \kappa_{g}-\partial_{n}\phi\right)  ds=0\text{\thinspace},
\end{equation}
and with $a\rightarrow b$ it follows that

\textbf{Condition 2:}

\textit{For a point }$a\in\Gamma_{j}\backslash\left(  J\cup P\right)
$\textit{ with boundary curvature }$\kappa_{g}$\textit{, }$\phi$\textit{ must
satisfy}
\[
\partial_{n}\phi=\kappa_{g}\text{.}
\]

The equation in Condition 2 is the same as Eq. (\ref{phi_manifold_geodesic}).
This is not incidental: Condition 2 causes cross-field geodesics, that are
close and parallel to the boundary, to follow the shape of the boundary, as
shown in Fig. \ref{boundary_aligment}.

For a point $c\in\left(  J\cup P\right)  $, that is, a junction or cone point
of the boundary, let $a,b\in\Gamma_{j}\backslash\left(  J\cup P\right)  $ be
two points on both sides of the junction point $c$, and $\beta$ the boundary
segment from $b$ to $a$, see Fig. \ref{condition3_boundary_singularity}. Let
$\alpha$ be a curve from $a$ to $b$. The Gauss-Bonnet theorem, Eq.
(\ref{gauss_bonnet}), for the curve $[\alpha,\beta]$ reads
\begin{equation}
\int_{\alpha}\kappa_{g}ds=2\pi-\int_{\beta}\kappa_{g}ds-\theta_{a}+\theta
_{b}-\left(  \pi-\theta_{in}\right)  -\int\int_{S}Kda,
\label{eq:cond_3_proof_1}
\end{equation}
where $\theta_{in}$ is inner angle at $c$, and $\theta_{a},\theta_{b}$ are
defined as in Eq. (\ref{eq:general_boundary_cond_equation}). Eq.
(\ref{eq:general_boundary_cond_equation}) reads

\begin{equation}
\theta_{b}-\theta_{a}=\int_{\alpha}\left(  \kappa_{g}-\partial_{n}\phi\right)
ds+n\frac{\pi}{2}. \label{eq:cond_3_proof_2}
\end{equation}
Adding Eq. (\ref{eq:cond_3_proof_1}) and Eq. (\ref{eq:cond_3_proof_2}) and
rearranging:
\[
\int_{\alpha}\partial_{n}\phi ds=n^{\prime}\frac{\pi}{2}+\theta_{in}
-\int_{\beta}\kappa_{g}ds-\int\int_{S}Kda.
\]
When $a\rightarrow c,b\rightarrow c$ then $\int_{\beta}\kappa_{g}
ds\rightarrow0,\int\int_{S}Kda\rightarrow0$, leading to:

\begin{figure}
[ptb]
\begin{center}
\includegraphics[
height=1.8893in,
width=1.7429in
]
{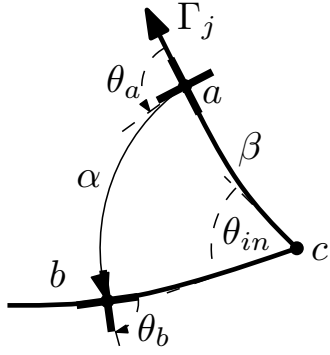}
\caption{Condition 3.}
\label{condition3_boundary_singularity}
\end{center}
\end{figure}

\textbf{Condition 3:}

\textit{For a curve }$\alpha$\textit{ as described above:}
\begin{equation}
\int_{\alpha}\partial_{n}\phi ds=n\frac{\pi}{2}+\theta_{in},
\label{eq:condition3}
\end{equation}
with $n\in
\mathbb{Z}
$. This requires $\partial_{n}\phi$ to have a singularity at $c$. On the
plane, for example, $\phi$ has a singularity of type $\frac{\partial\phi
}{\partial r}\sim1/r\,$\ at distance $r$ from $c$.

The final condition to be imposed on $\phi\,$\ is a relation between different
boundary components, and accordingly, its formulation is not local. It is just
a restatment of Eq. (\ref{eq:general_boundary_cond_equation}).

\textbf{Condition 4:}

\textit{For two boundary components }$\Gamma_{1},\Gamma_{2}\,$\textit{, and a
curve }$\alpha$\textit{ connecting }$a_{1}\in\Gamma_{1},$\textit{ to }
$a_{2}\in\Gamma_{2}$\textit{ it is necessary that}
\begin{equation}
\int_{a_{1}}^{a_{2}}\partial_{n}\phi ds=\theta_{a_{2}}-\theta_{a_{1}}
+\int_{a_{1}}^{a_{2}}\kappa_{g}ds+n\frac{\pi}{2}, \label{eq:condition_4}
\end{equation}
\textit{for an integer }$n$\textit{. }$\theta_{a_{1}},\theta_{a_{2}}$ are
defined as in Eq. (\ref{eq:general_boundary_cond_equation}).

Condition 4 states that the total flux through a curve connecting the two
boundary components can only belong to a certain, discrete, set of values.

\begin{figure}
[ptb]
\begin{center}
\includegraphics[
height=1.7763in,
width=1.9579in
]
{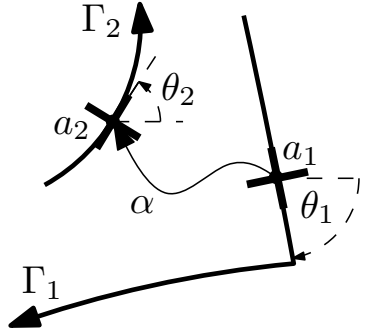}
\caption{Condition 4 for a planar domain.}
\label{condition4_2_boundaries_planar}
\end{center}
\end{figure}

Condition 4 has can be put in a simpler form when the domain is planar. Let
$\theta_{i}$ be the angle between the $x$-axis and $\mathbf{T}_{\Gamma_{1}
}\left(  a_{1}\right)  $, see Fig. \ref{condition4_2_boundaries_planar}. Then,
because $\int_{a_{1}}^{a_{2}}\kappa_{g}ds$ is equal to the change from $a_{1}$
to $a_{2}$ in the angle between the tangent to $\alpha$ and the $x$-axis we
have
\[
\theta_{a_{2}}-\theta_{a_{1}}+\int_{a_{1}}^{a_{2}}\kappa_{g}ds=\theta
_{2}-\theta_{1}.
\]
Condition 4 for the planar case then reads

\textbf{Condition 4 (planar domain):}

\textit{For two boundary components }$\Gamma_{1},\Gamma_{2}\,$\textit{, and a
curve }$\alpha$\textit{ connecting }$a_{1}\in\Gamma_{1},$\textit{ to }
$a_{2}\in\Gamma_{2}$\textit{ it is necessary that}
\[
\int_{a_{1}}^{a_{2}}\partial_{n}\phi ds=\theta_{2}-\theta_{1}+n\frac{\pi}{2},
\]
\textit{for an integer }$n$\textit{.}

\subsection{Sufficiency of the conditions \label{sec:sufficient_conditions}}

In the previous sections Conditions 1-4 were shown to be necessary for the
existance of a $\phi$-manifold. This section considers the question: when are
Conditions 1-4 sufficient? It turns out that the answer to this question
depends on the genus of the surface. Intuitively, the genus of a surface is
the number of handles a surface has \cite{hatcher}. For example a sphere or a
disk are genus-0 surfaces, and a torus is a genus-1 surface. A surface of any
genus can have any number of boundaries: for example, any planar domain, with
an arbitrary number of boundaries, is a genus-0 surface.

The following theorem states that for genus zero surfaces, Conditions 1-4 are
sufficient. The theorem also shows how to construct the cross-field given the
function $\phi$. For higher genus surfaces, an additional condition is
required, constraining the parallel transport along curves \textquotedblleft
passing through\textquotedblright\ the handles of higher genus surfaces. A
detailed disscusion is out of the scope of the present work, but a brief
account is given in Appendix \ref{appen:theorem_proof}.

\begin{theorem}
\label{theorem:sufficient} Suppose $D$ is a surface of genus-$0$. Let
$a_{0}\in\Gamma_{j_{0}}$ be some boundary point such that $a_{0}\notin J\cup
P$. Assume $\phi$ satisfies Conditions (1),(2),(3), and satisfies Condition
(4) with curves $\varepsilon_{i}$ between $a_{0}$ and points $e_{i}$
$\in\Gamma_{i}$, for every $i\neq j_{0}$. Let $V\left(  a_{0}\right)  $ be the
(unique) cross such that $\mathbf{T}_{\Gamma_{j_{0}}}\left(  a_{o}\right)  \in
V\left(  a_{0}\right)  $. For every $b\in D\backslash P$, let $\alpha$ be some
curve from $a_{0}$ to $b$, and define $V\left(  b\right)  $ to be the
translate of $V\left(  a_{0}\right)  $ along $\alpha$. Then $V\left(
b\right)  $ is independent of $\alpha$, and $V$ is a cross-field.
\end{theorem}

The proof is given in Appendix \ref{appen:theorem_proof}.

\begin{remark}
If the surface is closed, i.e. has no boundaries, the cross at some point
$a_{0}$ on the surface must be fixed for the cross-field to be unique.
\end{remark}

\section{Discussion: Finding a $\phi$-manifold\label{sec:discussion}}

According to Theorem \ref{theorem:sufficient} in
section\ \ref{sec:sufficient_conditions}, if a function $\phi$ is found,
satisfying Conditions 1-4, a $\phi$-manifold exists, and its cross-field can
be calculated.

In addition to Conditions 1-4, a meshing problem can include other
requirements, such as cell size requirements on the boundaries, or inside the
mesh. Clearly, finding an appropriate $\phi$-manifold then depends on these
conditions as well. In what follows the problem of finding a $\phi$-manifold
under different requirements will be discussed.

The next three sections focus on the planar case, and give concrete examples
of the theory presented above. The fourth section gives an example of a curved
surface meshing, that is solved analytically.

After finding a valid $\phi$-manifold, the final stage of a mesh generation
process involves finding a discrete partition into well-shaped manifold cells,
see below. A comprehensive discussion of this step will not be given here, but
some comments on this process will be made as part of the case studies.

\subsection{Case Study A: No singularities are needed\label{case_A}}

Sections \ref{case_A},\ref{sec:case_B},\ref{sec:case_C} deal with meshing of
planar domains. On the plane, the geodesic curvature $\kappa_{g}$ reduces to
the curvature of a planar curve, which will be denoted by $\kappa$.

We start with a simple planar case. Suppose that the boundary has only one
connectivity element (loop) $\Gamma_{1}$. Furthermore, suppose that all
junction angles $\theta_{J}{}_{i}$ (equal to $\pi-\theta_{in}$, $\theta_{in}$
the inner angle) are multiples of $\pi/2$, and that the their sum is $\sum
_{i}\theta_{J}{}_{i}=2\pi$. In such a case, Condition 4 is empty (since there
is only one boundary loop), and Condition 3 can be satisfied with $n_{i}
=-1\,$, and without any additional singularities at the junction points.
Condition 2 are Neumann boundary conditions on $\phi$. $\Gamma_{1}$ is a
simple loop, and the sum of junction angles is $2\pi$, thus the total is flux
of $\phi$ through the boundary is
\[
\Phi_{\Gamma_{1}}=\int_{\Gamma_{1}}\frac{\partial\phi}{\partial n}
ds=\int_{\Gamma_{1}}\kappa ds=2\pi-\sum_{i}\theta_{J}{}_{i}=0.
\]
Therefore given the boundary conditions a solution to the Laplace equation
(that is, with no singularities) exists, and is unique up to an additive
constant \cite{garabedian}.

\begin{figure}
[ptb]
\begin{center}
\includegraphics[
height=1.6855in,
width=2.7501in
]
{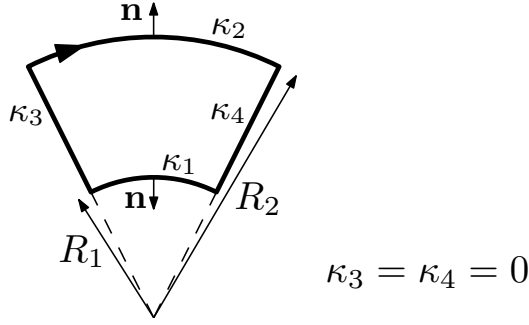}
\caption{The boundary specified in Eq. (\ref{eq: boundary conditions polar2}
).}
\label{polar_boundaries}
\end{center}
\end{figure}

A simple example of such a case, that can be solved analytically, is a section
of an annulus between two radial lines, see Fig. \ref{polar_boundaries}. The
boundary conditions on the sides of the boundary, dictated by the shape of the
boundary are given, according to Condition 2, by (note that $\kappa$ can be
negative, see Eq. (\ref{kappa_is_dtheta_ds})):
\begin{align}
\left.  \frac{\partial\phi}{\partial n}\right\vert _{1}  &  =\kappa_{1}
=\frac{1}{R_{1}}\label{eq: boundary conditions polar2}\\
\left.  \frac{\partial\phi}{\partial n}\right\vert _{2}  &  =\kappa_{2}
=-\frac{1}{R_{2}}\nonumber\\
\left.  \frac{\partial\phi}{\partial n}\right\vert _{3}  &  =\left.
\frac{\partial\phi}{\partial n}\right\vert _{4}=0.\nonumber
\end{align}

\begin{figure}
[ptb]
\begin{center}
\includegraphics[
height=1.638in,
width=5.0704in
]
{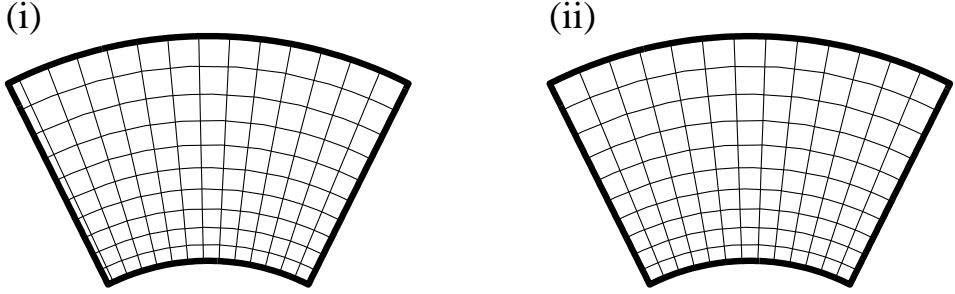}
\caption{C-frame geodesics for the example in Fig. \ref{polar_boundaries}. (i)
equally-spaced geodesics. (ii) geodescis forming well-shaped manifold cells.}
\label{polar2}
\end{center}
\end{figure}

The solution to the Laplace equation $\nabla^{2}\phi=0$ with boundary
conditions given in Eq. (\ref{eq: boundary conditions polar2}) is
\begin{equation}
\phi=-\ln r+C^{\prime}. \label{eq: polar solution}
\end{equation}
Here, $r$ is the distance from the center of the ring, and $C^{\prime}$ is a
constant. Note that no specification of the local cell size was given in the
input, and, according to section \ref{sec:relation_to_mesh_generation}, the
resulting local cell-size function is part of the solution. It is given by
$\exp\left(  -\phi\right)  =Cr$, with $C\equiv-\ln C^{\prime}$.

\begin{figure}
[ptb]
\begin{center}
\includegraphics[
height=1.6518in,
width=4.7668in
]
{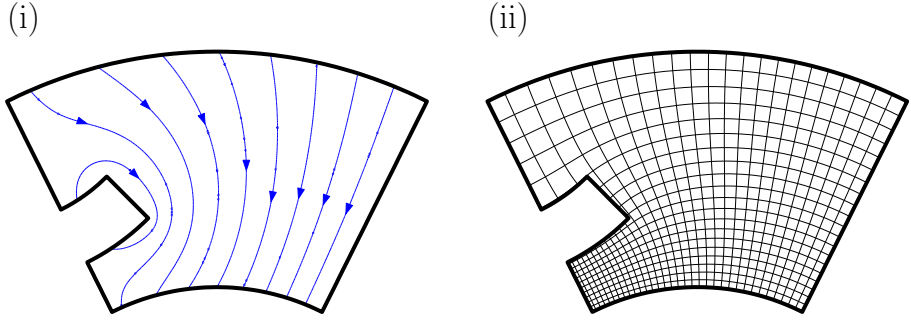}
\caption{A more elaborate exmaple than that of Eq.
(\ref{eq: boundary conditions polar2}). (i) Flow-lines of $\mathbf{\nabla}
\phi$. (ii) Equally-spaced geodesics.}
\label{polar_recess_geodesics_and_flow}
\end{center}
\end{figure}

\begin{figure}
[ptbh]
\begin{center}
\includegraphics[
height=2.252in,
width=4.9623in
]
{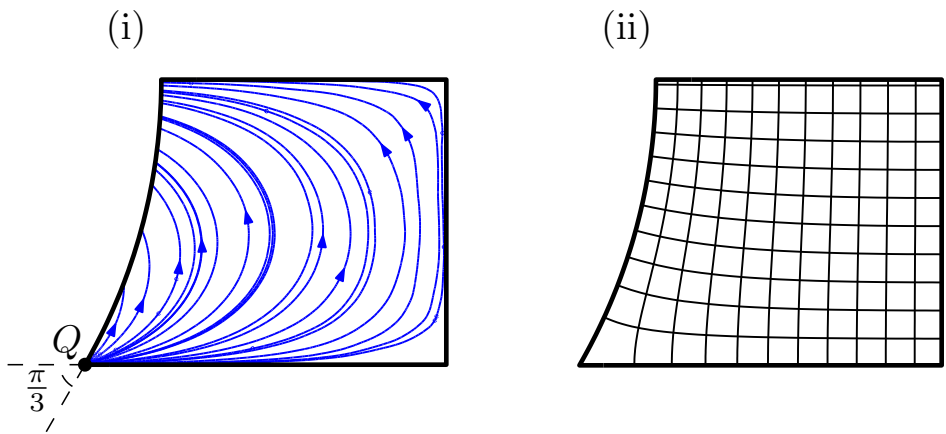}
\caption{A domain with a singularity at a junction point.}
\label{splash_flow_and_mesh}
\end{center}
\end{figure}

Fig. \ref{polar2} shows geodesics starting from the boundaries. To create the
figure, as well as Fig. (\ref{polar_recess_geodesics_and_flow}), Fig.
(\ref{splash_flow_and_mesh}) and Fig. (\ref{double_diamond_ignore_both}
),(iii), the Poisson equation was solved numerically on a triangular mesh
inside the domain (even though an analytical solution is known in the case
shown in Fig. (\ref{polar2})). The geodesics were calculated by solving the
geodesic equation, Eq. (\ref{phi_manifold_geodesic}), with initial tangent
perpendicular to the boundary. The\ integration constant $C$ was chosen such
that 10 manifold-edges will fit on the radial boundaries, i.e. that
$\int_{R_{1}}^{R_{2}}e^{\phi}dr=10$. This fixes the solution completely, and
gives a non-integer manifold-length along the arcs. Fig. \ref{polar2},(i)
shows geodesics spaced at unit manifold-distances of each other. Whereas the
radial direction fits exactly 10 manifold-edges of equal length, the
manifold-distance from the left-most geodesic to the boundary is less then 1,
resulting in manifold cells with high aspect ratio in the left-most row of
cells. In Fig. \ref{polar2}, (ii), two different spacings are used, so as to
allow equal spacing in both the radial and tangential directions, with an
aspect ratio as close to one as possible. Fig.
\ref{polar_recess_geodesics_and_flow} shows an example with a more elaborate
boundary adhering to the restrictions stated in the beginning of section
\ref{case_A}. Again, the equally-spaced geodesics of Fig.
\ref{polar_recess_geodesics_and_flow},(ii) do not form well-shaped (or even
valid) cells\footnote{The geodesics shown near the right and bottom of the
intrusion in Fig. \ref{polar_recess_geodesics_and_flow}\ are not parallel to
the boundary. This is due to the change in cell size. Other geodesics, closer
to the boundary, follow the shape of the boundary more closely.}. A valid
discretization can be created, e.g. by using geodesics emanating from the
junction points to decompose the domain.

\begin{remark}
Note that in general, local edge directions are \emph{not} aligned with the
flow lines of $\nabla\phi$, as can be seen e.g in Fig.
\ref{polar_recess_geodesics_and_flow}.
\end{remark}

We now lift the restriction that all junction angles are multiples of $\pi/2$.
If the angle is not $\pi/2$, Condition 3 requires that $\phi$ have a
radial-singularity at the junction point. Denote the junction inner-angle by
$\theta_{in}$. Suppose that $\phi$ contains a singularity caused by placing a
charge at the junction point, i.e. $\phi=\frac{Q}{2\pi}\ln r$. (If two or more
boundary segments are incident on the same point, other functions may be
required.) Let $\mathbf{\alpha}_{r}$ be the curve formed by traversing an arc
of the circle at a distance $r$ from the junction point in a counter-clockwise
direction, as in Fig. \ref{condition3_boundary_singularity}. Then according to
Condition 3 (Eq. (\ref{eq:condition3}))
\[
\frac{\pi}{2}n^{\prime}+\theta_{in}=\int_{\mathbf{\alpha}_{r}}\frac
{\partial\phi}{\partial n}ds=-\frac{Q}{2\pi r}\theta_{in}r=-\frac{Q}{2\pi
}\theta_{in},
\]
for $n^{\prime}\in
\mathbb{Z}
$, so
\begin{equation}
Q=2\pi\left(  n\frac{\pi/2}{\theta_{in}}-1\right)
\label{eq: charge_for_angle_sigularity}
\end{equation}
for $n=-n^{\prime}$. $n$ must be positive, since otherwise $Q\geq2\pi$,
causing the manifold-area to diverge at the singularity, by the same argument
as presented in section \ref{sec:cone_points}. Apart from this restriction the
number $n$ is not fixed, and affects the manifold-angle\ at the singularity,
i.e. the number of manifold-cells incident on the junction.

The example in Fig. \ref{splash_flow_and_mesh} shows a domain enclosed by one
boundary element. In one junction, the inner angle is $\pi/3$, which,
according to Eq. (\ref{eq: charge_for_angle_sigularity}), requires a
singularity of charge $Q=\pi$, for $k=1$ placed at the junction. The solution
was calculated by decomposing $\phi$ into 2 contributions: $\phi=\phi_{c}
+\phi_{L}$. $\phi_{c}$ is the charge potential, $\phi_{c}=\frac{Q}{2\pi}\ln
r$. $\phi_{L}$ was computed by numerically solving the Laplace equation on a
triangular mesh with boundary conditions $\frac{\partial\phi_{L}}{\partial
n}=$ $\frac{\partial\phi}{\partial n}-\frac{\partial\phi_{c}}{\partial n}$.

The examples that were presented in this case study could have been obtained
by a conformal mapping from a \textquotedblleft logical\textquotedblright
\ domain. This is, of course, not true when cone points are present inside the
domain\footnote{Even without cone-points inside the domain, this is not always
possible, since the mapping from $D$ to the \textquotedblleft
logical\textquotedblright\ domain is not, in general, one-to-one. In
manifold-theory terminology, even if the manifold is flat and simply
connected, it is not necessarily covered by a single geodesic coordinate patch
of the conformal metric.}.

Note, however, that unlike many mapping techniques, even if such a
\textquotedblleft logical\textquotedblright\ domain can be defined, its shape
\emph{is not fixed in advance, and is part of the solution}. This is even more
pronounced in problems involving cone points, see below. Conformal mappings
that are also boundary aligned are quite limited in the scope of problems they
can mesh, and sometimes yield large differences in cell size (as in the
example shown in Fig. \ref{polar_recess_geodesics_and_flow}). That is why in
mapping techniques the conformal restriction is lifted, see e.g. Ref. 2. In
the present work, the conformal condition (in its manifold formulation) is
retained, and instead cone points are allowed into the manifold.

\subsection{Case Study B:\ Two boundaries, no boundary cell-size
demand\label{sec:case_B}}

The purpose of the example in this section is to demonstrate the meaning and
relevance of Condition 4. Unlike the other boundary alignment conditions, the
formulation of Condition 4, stating the required relations between different
boundary elements, is not local. We now show that the relative placement of
different boundary elements can force the introduction of cone points in order
to obtain a valid cross-field.

\begin{figure}
[ptb]
\begin{center}
\includegraphics[
height=1.5912in,
width=4.8205in
]
{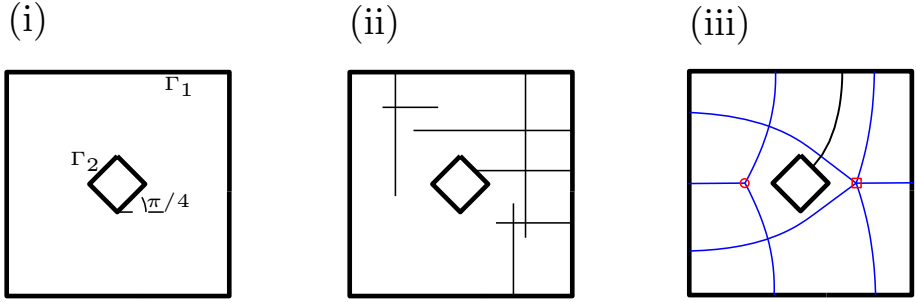}
\caption{The significance of Condition 4. (i) The input boundary. (ii)
Ignoring Condition 4, $\phi=const$ is a possible manifold. The resulting cross
field does not conform with all boundaries. (iii) A manifold with two
singularities. Selected cross field geodesics are drawn. }
\label{double_diamond_ignore_both}
\end{center}
\end{figure}

Fig. \ref{double_diamond_ignore_both},(i) shows a domain bounded by two
boundary elements, an \textquotedblleft outer loop\textquotedblright
\ $\Gamma_{1}$, and an \textquotedblleft inner loop\textquotedblright
\ $\Gamma_{2}$. Assume that there are no cell-size demands. We start by
ignoring Condition 4, and trying to proceed as in Case A, that is, searching
for a solution without any cone points. The curves composing $\Gamma
_{1},\Gamma_{2}$ are straight lines, so the Neumann boundary conditions read
$\frac{\partial\phi}{\partial n}=0$, and
the solution to the Laplace equation is trivial: $\phi=const$. Condition 3 is
also fulfilled with $k=1$ at all junctions. However, this solution does not
give a valid cross-field. A cross-field geodesic running from $\Gamma_{1}$ to
$\Gamma_{2}$ will not reach $\Gamma_{2}$ at a right angle, see Fig.
\ref{double_diamond_ignore_both},(ii). Thus, we cannot do without Condition
(4), and since the \emph{only} solution without cone points with boundary
conditions $\frac{\partial\phi}{\partial n}=0$ is $\phi=const$, we learn that
a solution without cone points does not exist. One possible solution with cone
points is shown in Fig. \ref{double_diamond_ignore_both},(iii). This solution,
obtained using symmetry arguments, contains two cone points with opposite
signs. Selected cross-field geodesics are shown in Fig.
\ref{double_diamond_ignore_both},(iii). One of them runs from one boundary to
the other. The rest are geodesics that are incident on the cone points.

\begin{remark}
\label{star_geodesics}In Fig. \ref{double_diamond_ignore_both},(iii),
cross-field geodesics reaching the cone points are shown. Three cross-field
geodesics reach the left cone point, which has a charge of $k=-1$, and five
reach the right cone point, which has a charge of $k=1$. Moreover, it seems
they reach the cone points at equally distributed angles. These observations
are true in general. It can be proved that there are always exactly $k+4$
angles from which cross-field geodesics are incident upon a singularity, and
these angles are equally distributed around the cone point. Such geodesics
will be called \emph{star-geodesics}. Note that this affects the angles of the
mesh-cells created: for small cells with a cone point on one vertex, that
vertex's inner angle will be approximately $\frac{2\pi}{k+4}$.
\end{remark}

\subsection{Case Study C: boundary cell-size demand; finding cone points'
locations\label{sec:case_C}}

In Case Studies A,B above no constraint on the size of boundary edges was
given. Yet the boundary edge-length is often specified in meshing problems.
Suppose we are given a function $F$ stating the required cell size (that is,
cell edge length) at each point on the boundary. The local cell size is
$e^{-\phi}$, thus $F=e^{-\phi}$, giving a Dirichlet boundary condition on
$\phi$:
\begin{equation}
\left.  \phi\right\vert _{\Gamma}=-\ln F. \label{eq: size_boundary_condition}
\end{equation}
Condition 1 states that $\phi$ obeys a Poisson equation with point charges
playing the role of cone points. The problem of finding a suitable $\phi$ is
reduced to finding a charge distribution (number of charges, their locations
and charge-strengths), such that $\phi$ will fulfill Conditions 2,3,4,
together with Eq. (\ref{eq: size_boundary_condition}). This is an
\textit{Inverse Poisson (IP) }problem . As opposed to a Direct Poisson
problem, where the charge-distribution $\rho$ in $\nabla^{2}\phi=\rho$ is
known, and one is asked to find $\phi$, in an IP problem,\emph{\ certain
information on }$\phi$\emph{\ is given, and the charge distribution }$\rho
$\emph{\ is to be found}.

IP problems have important applications in various areas of science and
engineering \cite{yamaguti}-\cite{zidarov}. By its nature, the IP problem is
ill-posed, and the solution may not be unique, and may be sensitive to small
changes of the input, such as small changes in boundary conditions. In
problems of this type any prior information on the charge distribution can
play an important role in the solution of the problem.

The problem of finding $\phi$ can be broken into the following steps:

\begin{description}
\item[(i)] Given the boundaries $\Gamma_{i}$ of the domain, and the cell-size
requirement $F$ on the boundary, calculate the Neumann boundary conditions
$\left.  \frac{\partial\phi}{\partial n}\right\vert _{\partial D}=\kappa$
(Condition (2)), and Dirichlet boundary condition $\left.  \phi\right\vert
_{\partial D}=$ $-\ln\left(  F\right)  $ (Eq.
(\ref{eq: size_boundary_condition})).

\item[(ii)] Impose boundary condition (3), e.g. by placing charges at the
junction points (see Eq. (\ref{eq: charge_for_angle_sigularity})).

\item[(iii)] Solve the IP problem: Find the (finite) number, location and
strength of charges such that Neumann and Dirichlet boundary conditions
calculated in (i), and Condition 4 hold approximately. According to Condition
1, the charges should be of strength $k_{i}\pi/2$, with $k_{i}>-4$. The
charges can be placed:

\begin{enumerate}
\item Inside $D$ (forming the set $P$).

\item At junction points (which amounts to changing $k$ in Eq.
(\ref{eq: charge_for_angle_sigularity})).

\item On the rest of the boundary (forming the set $P\cap\partial D$), where
according to Eq. (\ref{eq: charge_for_angle_sigularity}) with $\theta_{in}
=\pi$, charges of strength $k_{i}\pi$ can be placed.
\end{enumerate}

\item[(iv)] Once the charges are placed, $\phi$ is found by solving the
standard (Direct) Poisson problem.
\end{description}

\begin{remark}
\begin{description}

\item[(i)] Note that since this is an inverse problem, i.e. the charge
distribution is not fixed, both Neumann and Dirichlet boundary conditions can
be imposed together.

\item[(ii)] Though the limit on $k_{i}$ due to Condition (1) is $k_{i}>-4$,
charges should be of charge $k_{i}\geq-1$ in order to have convex cells, and
preferably with $k_{i}\leq2$. This is due to inner angles of cells incident on
the singularity, see Remark \ref{star_geodesics} in section \ref{sec:case_B}.
\end{description}
\end{remark}

\begin{figure}
[ptbh]
\begin{center}
\includegraphics[
height=2.949in,
width=3.8994in
]
{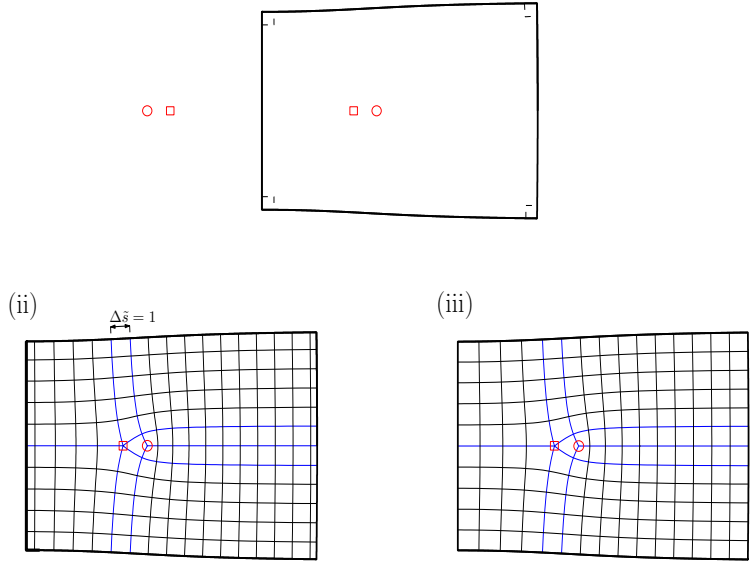}
\caption{The steps of a possible mesh-generation process. (i) The input
boundary (thick line). Locations of singularities of $\phi_{0}$ that was used
to \textit{create} the boundary are marked. (Square: $k=1$, Circle: $k=-1$).
(ii) A solution recovered using an IP algorithm (see text). Reconstructed
singularities marked as in (i). Equally-spaced geodesics are shown (thin
lines). (iii) Geodesics of the reconstructed solution, at approximately equal
spacings.}
\label{double_dislocation_mesh}
\end{center}
\end{figure}

\begin{figure}
[ptbh]
\begin{center}
\includegraphics[
height=1.8481in,
width=2.4682in
]
{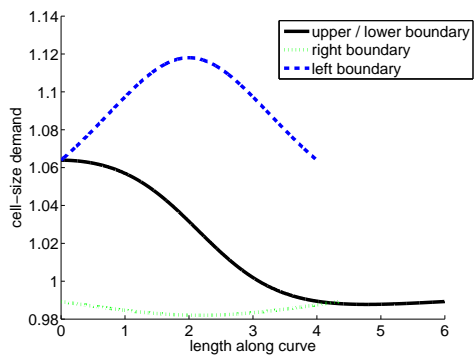}
\caption{The cell-size input requirement along the four sides of the boundary
shown in Fig. \ref{double_dislocation_mesh},(i). The upper / lower boundary
curves are traced from left to right.}
\label{edge_length_demand}
\end{center}
\end{figure}

The steps outlined above are illustrated in Fig. \ref{double_dislocation_mesh}
. For the purpose of this example, an input to the algorithm described above
was created artificially by joining four geodesics of a $\phi$-field chosen in
advance, at right angles to each other. The $\phi$-field chosen for creating
the boundary is a sum of fields from four charges: $\phi_{0}\left(  r\right)
=\sum_{i=1}^{4}k_{i}\frac{\pi}{2}\ln\left\vert r-r_{i}\right\vert $, with
$k_{i},r_{i}$ the charges' strengths and locations. Two charges are located
inside the boundary, and two outside, see Fig. \ref{double_dislocation_mesh}
,(i). This defines the shape of the boundary. The cell size requirement
imposed on the boundary was $F=e^{-\phi_{0}}$, plotted in Fig.
\ref{edge_length_demand}. The boundary shape and the cell size demand are the
input of the problem.

Using this input, the steps of the algorithm outlined above were followed.
First (step (i)), Neumann and Dirichlet boundary conditions were calculated
from the input. Step (ii) was fulfilled automatically without adding
additional charges, because all junction have right inner angles. In step
(iii) an algorithm solving the IP problem \cite{el-badia} was invoked. The
algorithm exactly reconstructed the location and charge of the two charges
inside the domain. (It is important to note that the IP algorithm of
\cite{el-badia} could be readily applied due to the artificial nature of the
example: the input was constructed such that these two charges will
reconstruct the input conditions exactly. This may not be the case in other
cases, and may require other IP solution methods.) The $\phi$-field was then
constructed by solving the \emph{direct}, standard Poisson problem, with
\emph{given} charges. In this example, since the location of the charges
inside the domain where recovered exactly, the $\phi$ recovered was exactly
$\phi_{0}$.

Fig. \ref{double_dislocation_mesh},(ii) shows cross-field geodesics at equal
manifold-distances $\Delta\tilde{s}=1$. As was discussed in section
\ref{sec:relation_to_mesh_generation}, the function $\phi$ is defined up to an
additive constant, that controls the overall cell-size. This constant was
chosen such that the manifold-distance $\Delta\tilde{s}$\ between two
geodesics incident on the two charges be equal to one, see Fig.
\ref{double_dislocation_mesh},(ii). In Fig. \ref{double_dislocation_mesh}
,(iii) cross-field geodesics spaced at approximately equal manifold-distances
are shown, such that no high aspect-ratio manifold cells, as those in Fig.
\ref{double_dislocation_mesh},(ii), exist. Note that the star-geodesics divide
the domain into sub-domains without charges, that can be meshed more easily.
This hints at possible ways of using the $\phi$-manifold for creating meshes.

\subsection{Other cases}

Other mesh requirements can be formulated. Examples include cell-size within
the domain (as in adaptive meshing), and cell direction. We comment briefly on
these subjects. In the case of adaptive meshing, a requirement for cell size
$F$ as a function of location inside the domain is given. This is translated
to a requirement on $\phi$ by $\phi=$ $-\ln\left(  F\right)  $. This fixes
$\phi$ completely, and therefore may not be fulfilled exactly along with other
conditions, such as having $\nabla^{2}\phi=0$ (in the planar case) almost
everywhere. However, weaker constraints may be given, such as specifying $F$
along curves within the domain. When specifing cell direction, the direction
of the cross-field is given. This translated to a constraint similar to
Condition 4, on the flux $\Phi$ through some arbitrary curve. As is the case
with adaptive meshing, a requirement for cell direction everywhere inside the
domain is too restrictive, but more limited demands may be applied, such as an
approximate alignment.

\subsection{A curved surface example}

For many surface meshing applications it is desired that the cells be aligned
with certain prescribed directions, such as the principle curvature directions
of the surface. As an example of a meshing problem of this type, the class of
surfaces known as surfaces of revolution is analyzed analytically in this
section, and it is shown how a cross-field aligned with the principle
directions is constructed in this case.
\begin{figure}
[ptb]
\begin{center}
\includegraphics[
height=2.0029in,
width=2.0029in
]
{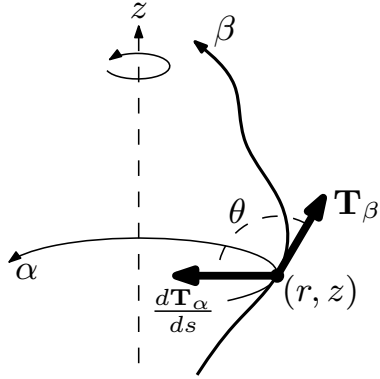}
\caption{Surface of revolution.}
\label{revolution_surface}
\end{center}
\end{figure}

A surface of revolution is defined by a curve in the $\left(  r,z\right)  $
plane, that is revolved around the $z$-axis, see Fig. \ref{revolution_surface}
. The direction of rotation will be called the $\theta$-direction. The line
traced by the curve at a given $\theta$ is known as a \emph{meridian}. The
circles at given $\left(  r,z\right)  $ values are known as \emph{circles of
revolution} \cite{millman}. The principle axis directions at a point are the
directions of the circle of revolution and meridian passing through that
point. A mesh aligned with these directions has $N$ cells around every circle
of revolution at a given $\left(  r,z\right)  $. The cell size at a given
point should therefore be $N/\left(  2\pi r\right)  $. But the cell size is
equal to $e^{-\phi}$, so $\phi$ is expected to be
\begin{equation}
\phi=-\ln\left(  \frac{N}{2\pi r}\right)  \text{.}
\label{eq:revolution_expected_cell_size}
\end{equation}

We now show that this is indeed the case. By the alignment requirement, edges
must be laid along circles of revolution and meridians. Therefore, these must
be geodesics of the manifold with metric $\tilde{g}_{ij}$. Let $\mathbf{T}
_{\alpha}\left(  s\right)  $ be the tangent to the curve $\alpha\left(
s\right)  $ tracing a circle of revolution at $\left(  r,z\right)  $, see Fig.
\ref{revolution_surface}. Then the geodesic curvature $\kappa_{g}\left(
s\right)  $ is equal to
\[
\kappa_{g}=\frac{d\mathbf{T}_{\alpha}}{ds}\cdot\mathbf{T}_{\beta}=\frac{1}
{r}\cos\theta
\]
where the angle $\theta$ between $\alpha$ and the meridian, $\beta$, is equal
to $\cos\theta=-\left(  \left(  dz/dr\right)  ^{2}+1\right)  ^{-1/2}$, so
\[
\kappa_{g}=-\frac{1}{r\sqrt{\left(  \frac{dz}{dr}\right)  ^{2}+1}}\text{.}
\]
For a geodesic of $\tilde{g}_{ij}$, $0=\tilde{\kappa}_{g}=\kappa_{g}
-\partial\phi/\partial n$. The normal is directed along the meridian, so
$\phi$ can is found by integrating along the meridian $\beta$. Noting that
$ds=\sqrt{\left(  \frac{dz}{dr}\right)  ^{2}+1}dr$ along $\beta$, we have
\[
\phi=\int\frac{\partial\phi}{\partial s}ds=\int-\frac{ds}{r\sqrt{\left(
\frac{dz}{dr}\right)  ^{2}+1}}=-\int\frac{1}{r}dr=-\ln\frac{C}{r}\text{.}
\]
As expected in Eq. (\ref{eq:revolution_expected_cell_size}). The constant $C$
is determined once for the surface, e.g. by the number of cells on a given
circle of revolution. By comparing with Eq.
(\ref{eq:revolution_expected_cell_size}), $C$ is interpreted as $C=N/2\pi$.

If the surface of revolution reaches the $z$-axis at some point $p$, and is
not parallel to the $z$-axis there, the manifold will have a cone points of
charge $-4\frac{\pi}{2}$ at $p$. To see this, we again use the fact that for a
circle of revolution $0=\tilde{\kappa}_{g}=\kappa_{g}-\partial\phi/\partial
n$. Integrating over a revolution circle surrounding $p$:
\begin{align*}
0  &  =\int\left(  \kappa_{g}-\frac{\partial\phi}{\partial n}\right)
ds=2\pi-\int\int Kda+\int\int\nabla^{2}\phi da\\
&  =2\pi+\int\int n\frac{\pi}{2}\delta_{p}^{\left(  2\right)  }da\text{.}
\end{align*}
Eq. (\ref{gauss_bonnet}),(\ref{eq:condition1}) were used. Thus, by $\int
\int\delta_{p}^{\left(  2\right)  }da=1$ we have $n=-4$. We note that in the
vicinity of a singularity of charge $-4\frac{\pi}{2}$ the number of cells
diverges, see section \ref{sec:cone_points}. If this is undesirable, a mesh
that isn't exactly aligned with the principal directions may be constructed,
e.g. by replacing the single charge with a few cone points with smaller charges.

\section{Conclusions}

In this work a continuum description of unstructured meshes was proposed.The
structure aims at describing meshes that, away from irregular vertices, have
well-shaped cells. In the limit of an increasingly finer mesh, the cell's
shape approaches the shape of a square (in the case of a quadrilateral mesh)
or equilateral triangle (for triangle meshes). Accordingly, in the continum
limit such meshes can be described by just two local properties: the local
cell size, and the local directions of edges, formalized by the notion of a cross.

The connection between cell size and cell direction is established by defining
a Riemmanian manifold, the $\phi$-manifold. The geodesics of the manifold
trace the edges of the mesh, as is formalized via the definition of a
cross-field. This analysis allows the focus to turn to the irregular vertices,
represented in the continuum structure by cone points.

The demand that the mesh conform to the boundary, and have a finite number of
cells, produces conditions on the function $\phi$. The resulting reduced
problem is an Inverse Poisson problem, of finding a distribution of localized
charges adhering to these conditions. The charges correspond to cone points.

The main component needed to apply the theory to mesh generation of planar
domains is a suitable Inverse Poisson algorithm. An algorithm for creating the
final discrete mesh is also required.

\section{Acknowledgements}

I am indebted to Dov Levine for introducing me to the theory of defects, from
which this work began. Helpful discussions with Mirela Ben-Chen, Michael
Entov, Craig Gotsman and Shlomi Hillel are greatly appreciated.

\appendix

\section{Relation between geodesic curvature at different
metrics\label{app:k_g_different_metrics}}

In this appendix Eq. (\ref{kappa_is_dtheta_ds}) is derived. Eq.
(\ref{kappa_is_dtheta_ds}) is a known relation in conformal geometry, however
sign and direction convensions vary, so a derivation is given for
completeness. Let $\left(  x^{1},x^{2}\right)  $ be a coordinate system,
$g_{ik}$ the metric components, and $g=\det\left(  g_{ij}\right)  $. For a
curve $\alpha$ parametrized by the length parameter $\alpha\left(  s\right)
=\left(  \alpha^{1}\left(  s\right)  ,\alpha^{2}\left(  s\right)  \right)  $,
$\kappa_{g}$ is given by \cite{laugwitz}
\begin{equation}
\kappa_{g}=\varepsilon_{li}\frac{d\alpha^{l}}{ds}\left(  \frac{d^{2}\alpha
^{i}}{ds^{2}}+\Gamma_{jk}^{\ \ i}\frac{d\alpha^{j}}{ds}\frac{d\alpha^{k}}
{ds}\right)  ,
\end{equation}
where $\varepsilon_{li}$ satisfies $\varepsilon_{11}=\varepsilon_{22}=0$,
$\varepsilon_{12}=-\varepsilon_{21}=\sqrt{g}$. $\Gamma_{jk}^{\ \ i}$ are the
Christoffel symbols, related to the metric by the formulas
\begin{equation}
\Gamma_{ij}^{\ \ l}=\frac{1}{2}g^{kl}\left(  \frac{\partial g_{ik}}{\partial
x^{j}}-\frac{\partial g_{ij}}{\partial x^{k}}+\frac{\partial g_{kj}}{\partial
x^{i}}\right)  , \label{eq:christofell_gij}
\end{equation}
where $g^{kl}$ is the $k,l$ entry of the inverse of the matrix $\left(
g_{ij}\right)  $.

Denote by $g_{ij}$ the \textquotedblleft standard\textquotedblright\ metric
induced by the embedding in three dimensional space, by $\tilde{g}_{ik}$ the
metric of the $\phi$-manifold, (see Definition \ref{def:phi_manifold}), and
the corresponding Christoffel symbols by $\Gamma_{ij}^{\ \ l},\tilde{\Gamma
}_{ij}^{\ \ l}$ respectively. Substituting $\tilde{g}_{ij}=e^{2\phi}g_{ij}$
(see Definition \ref{def:phi_manifold}) into Eq. (\ref{eq:christofell_gij}),
the equation for $\tilde{\Gamma}_{ij}^{\ \ l}$\ reads
\begin{equation}
\tilde{\Gamma}_{ij}^{\ \ l}=\Gamma_{ij}^{\ \ l}+\phi,_{j}\delta_{i}^{l}
-\phi,_{k}g^{kl}g_{ij}+\phi,_{i}\delta_{j}^{l}\ \ \ \text{,}
\label{eq:christ_relation}
\end{equation}
where comas denote differentiation: $\phi,_{i}\equiv\partial\phi/\partial
x^{i}$. Eq. (\ref{eq:christ_relation}) can now be used to derive a relation
between $\kappa_{g}$ and $\tilde{\kappa}_{g}$, the geodesic curvatures in the
two metrics, at a point $p$ on the curve $\alpha$. The derivation is simpler
if a normal coordinate system is chosen, for which $g_{ij}\left(  p\right)
=\delta_{ij}$, and $\Gamma_{ij}^{\ \ l}\left(  p\right)  =0$ (this can always
be done, see \cite{laugwitz}). Noting that $\sqrt{\tilde{g}}=e^{2\phi}\sqrt
{g}$, $d\tilde{s}=e^{\phi}ds$, the equation for $\tilde{\kappa}_{g}$ reads
\begin{align*}
\tilde{\kappa}_{g}  &  =\tilde{\varepsilon}_{li}\frac{d\alpha^{l}}{d\tilde{s}
}\left(  \frac{d^{2}\alpha^{i}}{d\tilde{s}^{2}}+\tilde{\Gamma}_{jk}
^{\ \ i}\frac{d\alpha^{j}}{d\tilde{s}}\frac{d\alpha^{k}}{d\tilde{s}}\right) \\
&  =e^{-\phi}\varepsilon_{li}\frac{d\alpha^{l}}{ds}\left(  \frac{d^{2}
\alpha^{i}}{ds^{2}}+\left(  \phi,_{k}\delta_{j}^{i}-\phi,_{m}g^{mi}g_{jk}
+\phi,_{j}\delta_{k}^{i}\right)  \frac{d\alpha^{j}}{ds}\frac{d\alpha^{k}}
{ds}\right) \\
&  =e^{-\phi}\left(  \kappa_{g}+2\phi,_{k}\frac{d\alpha^{k}}{ds}\left(
\frac{d\alpha^{l}}{ds}\frac{d\alpha^{i}}{ds}\varepsilon_{li}\right)
-\frac{d\alpha^{l}}{ds}\varepsilon_{li}\delta^{im}\phi,_{m}\left(
\frac{d\alpha^{j}}{ds}\frac{d\alpha_{j}}{ds}\right)  \right) \\
&  =e^{-\phi}\left(  \kappa_{g}-\partial_{n}\phi\right)  .
\end{align*}
The last equality holds because $\frac{d\alpha^{l}}{ds}\frac{d\alpha^{i}}
{ds}\varepsilon_{li}=0\,$, as can be directly verified, because $\frac
{d\alpha^{j}}{ds}\frac{d\alpha_{j}}{ds}=1$ in arc length parametrization. The
convension that $\left(  \mathbf{T}_{\alpha},\mathbf{N}_{\alpha}\right)  $
form a right-hand system has been used.

\section{Triangular Meshes\label{Triangular_mesh}}

In this appendix the changes in the definitions and results in sections
\ref{sec:cross_field},\ref{sec:relation_to_mesh_generation}, and
\ref{sec:cone_points} for the case of triangular meshes are outlined.

\begin{definition}
[triangle cross field]Define an equivalence $\sim$ of vectors in
$\mathbb{R}^{2}$: for 2 vectors $\mathbf{v}_{1},\mathbf{v}_{2}\in\mathbb{R}^{2}$, $\mathbf{v}_{1}\sim\mathbf{v}_{2}$ if and only if $\mathbf{v}_{1}$,$\mathbf{v}_{2}$ are form an angle of $n\pi/3$, with $n\in\mathbb{Z}$. A \emph{cross} is an element of $\mathbb{R}^{2}/\sim$.
\end{definition}

The cross-field and $\phi$-manifold are the same as for the quadrilateral case.

The conditions for the existence of a triangle cross-field are:\newline
\newline Condition 1 for triangular meshes reads:\newline\newline
\textbf{Condition 1. }$\phi$\textit{\ obeys the equation}
\[
\nabla^{2}\phi\left(  r\right)  =K+\frac{\pi}{3}\sum_{i=1..N}k_{i}
\delta_{p_{i}}^{\left(  2\right)  },
\]
\textit{the Poisson equation with point sources (delta functions) }
$\delta_{p_{i}}^{\left(  2\right)  }$\textit{, with }$k_{i}\in
\mathbb{Z}
$\textit{, }$k_{i}>-6$\textit{.}\newline\textbf{Condition 2.} Is the same as
in the quadrilateral case.\newline\newline\textbf{Condition 3.} Eq.
(\ref{eq:condition3}) for the triangular case reads
\[
\int_{\alpha}\partial_{n}\phi ds=n\frac{\pi}{3}+\theta_{c}.
\]
\newline\newline\textbf{Condition 4.} Eq. (\ref{eq:condition_4}) of becomes
\[
\int_{a_{1}}^{a_{2}}\partial_{n}\phi ds=\theta_{a_{2}}-\theta_{a_{1}}
+\int_{a_{1}}^{a_{2}}\kappa_{g}ds+n\frac{\pi}{3}.
\]

\begin{figure}
[ptb]
\begin{center}
\includegraphics[
height=1.8057in,
width=3.1877in
]
{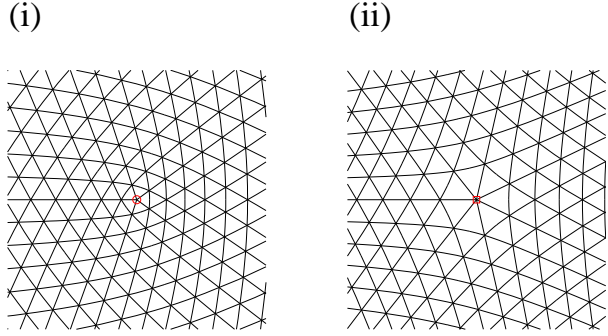}
\caption{Triangle cross field geodesics around a singularity. The geodesics
are spaced at unit manifold-distances. (i) A $k=-1$ singularity. (ii) A $k=1$
singularity.}
\label{defects_tri_5_7}
\end{center}
\end{figure}

Fig. \ref{defects_tri_5_7} shows selected triangle cross-field geodesics
around a singularity (with $\phi_{L}=0$), spaced at manifold-distances of
$\Delta\tilde{s}=1$\ from each other, for different singularity strengths.

\section{Higher-genus surfaces, and a proof of Theorem
\ref{theorem:sufficient} \label{appen:theorem_proof}}

This appendix includes a proof of Theorem \ref{theorem:sufficient}, but first,
the case of surfaces with genus higher than zero is briefly disscussed. Basic
notions of algebraic-topology are used, see e.g. \cite{hatcher}.

As the proof of Theorem \ref{theorem:sufficient} presented below shows, for a
cross-field to exist, the angle change in the parallel-transport of a vector
along a loop based at $a_{0}$ must be a multiple of $\pi/2$. Let $D^{\prime}$ be
the surface $D$ with holes are cut around the cone points of $D$. Let $a_{0}$
be a base point, as in Theorem \ref{theorem:sufficient}. Let $\left\{
\alpha_{i}\right\}  $ be a set of loops starting at $a_{0}$, each encircling a
single boundary of $D^{\prime}$ (including the holes cut around the cone
points). The angle change due to parallel transport around these loops is a
multiple of \ $\pi/2$, see the proof of the theorem below. Complete the set
$\left\{  \alpha_{i}\right\}  $ to a homotopy basis \cite{hatcher} of $D^{\prime}$, by adding
a set of loops $\left\{  \beta_{i}\right\}  $ based at $a_{0}$. The additional
condition is that the angle change due to parallel transport around a loop of
$\left\{  \beta_{i}\right\}  $ is a multiple of \ $\pi/2$. Any loop based at
$a_{0}$ is homotopic to a composition of loops from the homotopy basis. Since
the manifold $D^{\prime}$ is flat with the conformal metric, parallel
transport is preserved by the homotopy, hence the change in the angle along
any loop is a multiple of $\pi/2$ and the cross-field is well-defined. The
second part of the proof below remains unchanged.

We now turn to the proof of Theorem \ref{theorem:sufficient}:

\begin{proof}
We first prove that $V\left(  b\right)  $ for some $b\in D\backslash P$ is
independent of $\alpha$. Let $\alpha_{1},\alpha_{2}$ be two curves from
$a_{0}$ to $b$. Denote $\mathbf{T}_{0}\equiv\mathbf{T}_{\Gamma_{j_{0}}}\left(
a_{o}\right)  $. We need to show that the parallel transport of $\mathbf{T}
_{0}$ to $b$ gives vectors that belong to the same cross, i.e.
\begin{equation}
\measuredangle\left(  \widetilde{PT}_{\mathbf{\alpha}_{2}}\mathbf{T}
_{0},\widetilde{PT}_{\mathbf{\alpha}_{1}}\mathbf{T}_{0}\right)  =k\frac{\pi
}{2} \label{eq: two parallel transports of t in 1-4 proof}
\end{equation}
for some $k\in
\mathbb{Z}
$. Using Eq. (\ref{kappa_is_dtheta_ds}),(\ref{kappa_g_relation}), Eq.
(\ref{eq: two parallel transports of t in 1-4 proof}) becomes
\begin{equation}
k\frac{\pi}{2}=\int_{\alpha_{1}}\left(  \kappa_{g}-\frac{\partial\phi
}{\partial n}\right)  ds-\int_{\alpha_{2}}\left(  \kappa_{g}-\frac
{\partial\phi}{\partial n}\right)  ds=\oint_{\alpha}\left(  \kappa_{g}
-\frac{\partial\phi}{\partial n}\right)  ds
\label{eq: 1 need to show in theorem}
\end{equation}
where $\alpha\equiv\lbrack\alpha_{1},\alpha_{2}^{-}]$. (Note that Eq.
(\ref{eq: 1 need to show in theorem}) is equivalent to the requirement that
the cross in $\alpha_{0}$ be parallel-translated to itself along $\alpha$.) We
now prove Eq. (\ref{eq: 1 need to show in theorem}). The region enclosed by
$\alpha$ can contain singularities and boundary curves. Surround them by
additional curves, as shown in Fig. \ref{sufficient_theorem1}. Denote the
region between the $\alpha$ curve and the $\omega$ and $\psi$ curves as $S$
(here the assumption that $D$ has genus zero enters). Since $\nabla^{2}\phi
=K$, Eq. (\ref{lap_phi_eq_K}), everywhere in $S$, and $\frac{\partial\phi
}{\partial n},\phi$ are finite on the boundary of $S$ (due to $\phi$-manifold
definition, (iii)), Green's theorem, Eq. (\ref{greens_theorem}), can be
applied in this case, giving
\begin{equation}
-\int_{S}Kda=-\int_{S}\nabla^{2}\phi da=\oint_{\alpha}\frac{\partial\phi
}{\partial n}ds+\sum_{i=1..N_{\omega}}\oint_{\omega_{i}}\frac{\partial\phi
}{\partial n}ds+\sum_{j=1..N_{\psi}}\oint_{\psi_{j}}\frac{\partial\phi
}{\partial n}ds \label{eq: total flux in theorem}
\end{equation}
where $N_{\omega}$ are the number of points in $P$, and $N_{\psi}$
inner-curves enclosed by $\alpha$. Consider a single loop $\psi_{r}$ around an
inner curve $\Gamma_{r}$, $1\leq r\leq N_{\psi}$. $\psi_{r}$ is composed of
curves $\eta_{i}$ around points of $J\cup P$ on the boundary, and curves
$\zeta_{i}$ between junctions, i.e. $\psi_{r}=[\zeta_{1},\eta_{1},\zeta
_{2},\eta_{2},...]$, see Fig. \ref{sufficient_theorem1}.
\begin{figure}
[ptb]
\begin{center}
\includegraphics[
height=3.122in,
width=4.4521in
]
{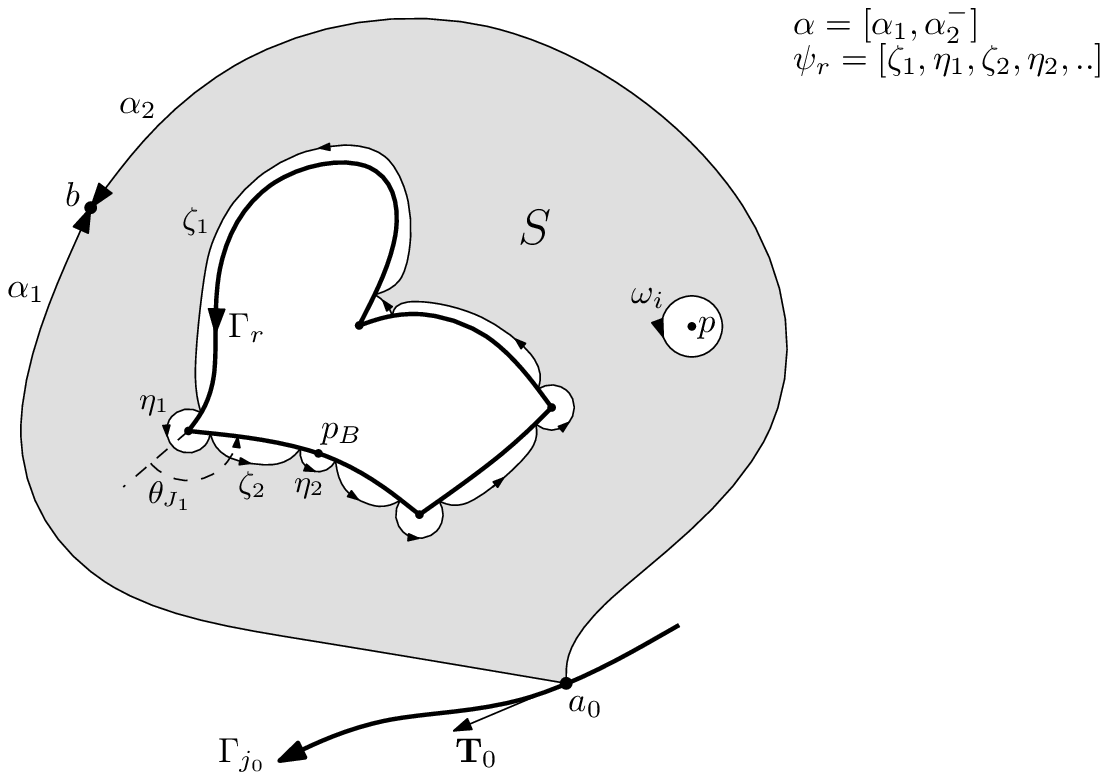}
\caption{Proving that the cross-field is well defined. $\Gamma_{r}$ is a
boundary curve. $p$ is a cone point. $p_{B}$ is a cone point on the boundary.}
\label{sufficient_theorem1}
\end{center}
\end{figure}
The total flux through $\psi_{r}$ is given by
\begin{align}
\oint_{\psi_{r}}\frac{\partial\phi}{\partial n}ds  &  =\sum_{i}\left(
\oint_{\mathbf{\eta}_{i}}\frac{\partial\phi}{\partial n}ds+\oint_{\zeta_{i}
}\frac{\partial\phi}{\partial n}ds\right)  =\sum_{i}\left(  \theta_{in_{i}
}+k_{J_{i}}\frac{\pi}{2}+\oint_{\zeta_{i}}\kappa_{g}ds\right)  =\nonumber\\
&  =\sum_{i}\left(  \left(  \pi-\theta_{J_{i}}\right)  +k_{J_{i}}\frac{\pi}
{2}+\oint_{\zeta_{i}}\kappa_{g}ds\right) \nonumber\\
&  =\sum_{i}\left(  \theta_{J_{i}}+\oint_{\zeta_{i}}\kappa_{g}ds+k_{\psi_{r}
}\frac{\pi}{2}\right)  \label{eq:total_flux_in_theorem}
\end{align}
with $k_{J_{i}},k_{\psi_{r}}\in
\mathbb{Z}
$. The second equality uses Conditions 2,3. The flux through $\omega_{i}$ is
given by Condition 1:
\begin{equation}
\oint_{\omega_{i}}\frac{\partial\phi}{\partial n}ds=k_{\omega_{i}}\frac{\pi
}{2}. \label{eq: total flux singularity}
\end{equation}
After substituting Eq. (\ref{eq:total_flux_in_theorem}
),(\ref{eq: total flux singularity}), Eq. (\ref{eq: total flux in theorem})
becomes
\begin{align*}
-\int_{S}Kda  &  =\oint_{\alpha}\frac{\partial\phi}{\partial n}ds+\sum
_{i}\left(  \theta_{J_{i}}+\oint_{\zeta_{i}}\kappa_{g}ds\right)  +\\
&  \sum_{i=1..N_{\omega}}k_{\omega_{i}}\frac{\pi}{2}+\sum_{i=1..N_{\psi}
}k_{\psi_{i}}\frac{\pi}{2},
\end{align*}
and using Gauss-Bonnet theorem for a domain with more then one boundary
component we find
\[
-\oint_{\alpha}\kappa_{g}ds-\oint_{\alpha}\frac{\partial\phi}{\partial
n}ds=k\frac{\pi}{2}
\]
for some $k\in
\mathbb{Z}
$. This proves Eq. (\ref{eq: 1 need to show in theorem}).

We now turn to prove the property (ii) of a cross-field, its boundary
alignment. Let $c$ be a boundary point, $c\in\Gamma_{r}\backslash\left(  J\cup
P\right)  $, $1\leq r\leq N_{\psi}$. As was shown above, the cross at $c$,
$V\left(  c\right)  $, is well-defined. It is left to show that $\mathbf{T}
_{\Gamma_{i}}\left(  c\right)  $ $\in V\left(  c\right)  $. Denote
$\mathbf{T}_{r}\equiv\mathbf{T}_{\Gamma_{r}}\left(  e_{r}\right)  $. The
assumptions of the theorem, together with Condition 4, assure that
$\mathbf{T}_{r}$ $\in V\left(  e_{r}\right)  $. Let $\beta$\textbf{ }be the
section of $\Gamma_{r}$ between $e_{r}$ and $c$. Define a curve $\alpha$ from
$e_{r}$ to $c$ composed of smooth curves $\eta_{n}$ around the points of
$J\cup P$ in the image of $\beta$, and $\mu_{m}$ curves between junctions, see
Fig. \ref{sufficient_theorem2}. The curves $\eta_{n}$ avoid the sigularities
that the function $\phi$ might have at points in $J\cup P$. Let $\theta
_{J_{n}}$ be the junction angle at the point of $J\cup P$ \textquotedblleft
bypassed\textquotedblright\ by the curve $\eta_{n}$ (zero for points that are
not juction points). Then, if $\eta_{n}$ follows $\Gamma_{r}$ closely, the
total turn of $\eta_{n}$ is equal to $\theta_{J_{n}}$:
\[
\int_{\eta_{n}}\kappa_{g}ds=\theta_{J_{n}}\text{,}
\]
as can be formally verified by applying the Gauss-Bonet theorem to the area
between $\eta_{i}$ and $\Gamma_{r}$. Note furthermore that $\theta_{J_{n}}
+\pi=\theta_{in_{n}}$, the inner angle at that point.
\begin{figure}
[ptb]
\begin{center}
\includegraphics[
height=2.5097in,
width=2.7095in
]
{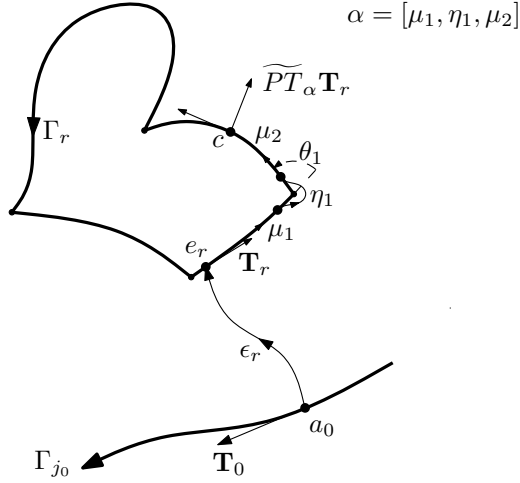}
\caption{Proving that the cross-field is aligned with the boundary.}
\label{sufficient_theorem2}
\end{center}
\end{figure}

The variation of $\mathbf{T}_{\Gamma_{r}}\left(  e_{r}\right)  $ along
$\alpha$ is
\begin{align*}
\measuredangle\left(  \widetilde{PT}_{\alpha}\mathbf{T}_{r},\mathbf{T}
_{\Gamma_{r}}\left(  c\right)  \right)   &  =\int_{\alpha}\tilde{\kappa}
_{g}d\tilde{s}=\int_{\alpha}\left(  \kappa_{g}-\partial_{n}\phi\right)  ds\\
&  =\sum_{m}\int_{\mu_{m}}\left(  \kappa_{g}-\partial_{n}\phi\right)
ds+\sum_{j}\int_{\eta_{j}}\kappa_{g}ds-\sum_{j}\int_{\eta_{j}}\partial_{n}\phi
ds\\
&  =\sum_{m}\int_{\mu_{m}}\left(  \kappa_{g}-\partial_{n}\phi\right)
ds+\sum_{j}\left(  \theta_{in_{j}}-\pi\right)  -\sum_{j}\int_{\eta_{j}
}\partial_{n}\phi ds\\
&  =\sum_{j}k_{j}\frac{\pi}{2}=k\pi/2
\end{align*}
with $k\in\mathbb{Z}$. Conditions 2 and 3 were applied for the $\mu_{m}$ and
$\eta_{j}$ curves, respectively. Thus $\mathbf{T}_{\Gamma_{i}}\left(
c\right)  \in$ $V\left(  C\right)  $, which completes the proof of the
boundary properties of the cross-field $V$.
\end{proof}

\end{document}